\documentclass[twocolumn,notitlepage,prb,superscriptaddress,longbibliography]{revtex4-2}
\usepackage{amsfonts}
\usepackage{textcomp}
\usepackage{times}
\usepackage{graphicx}
\usepackage{float}
\usepackage{latexsym,amsmath,amssymb,bm,euscript}
\usepackage[dvipsnames]{xcolor}
\usepackage{subfigure}
\usepackage{epstopdf}
\usepackage[colorlinks=true,linkcolor=blue,citecolor=blue,urlcolor=blue]{hyperref}
\usepackage{hyperref}
\usepackage{soul}
\usepackage[normalem]{ulem}
\usepackage{mathrsfs}
\usepackage{amsmath}
\usepackage{amstext}
\usepackage[bottom]{footmisc}
\usepackage{amsbsy}
\usepackage{ulem}
\usepackage{color}

\begin{document}

\title{Sign structure of the $t$-$t^\prime$-$J$ model and its physical consequences}
\author{Xin Lu}\thanks{Both authors contributed equally to this work.}
\affiliation{School of Physics, Beihang University, Beijing 100191, China}

\author{Jia-Xin Zhang}\thanks{Both authors contributed equally to this work.}
\affiliation{Institute for Advanced Study, Tsinghua University, Beijing 100084, China}

\author{Shou-Shu Gong}
\email{shoushu.gong@gbu.edu.cn}
\affiliation{School of Physical Sciences, Great Bay University, Dongguan 523000, China}
\affiliation{Great Bay Institute for Advanced Study, Dongguan 523000, China}

\author{D. N. Sheng}  
\email{donna.sheng1@csun.edu}
\affiliation{Department of Physics and Astronomy, California State University, Northridge, California 91330, USA}

\author{Zheng-Yu Weng}
\email{weng@mail.tsinghua.edu.cn}
\affiliation{Institute for Advanced Study, Tsinghua University, Beijing 100084, China}

\begin{abstract}
Understanding the doped Mott insulator is a central challenge in condensed matter physics. In this work, we first explicitly identify a new sign structure in the $t$-$t'$-$J$ model on the square lattice that replaces the conventional Fermi statistics for weakly interacting electrons. Then we show that the singular, i.e., the phase-string part of the sign structure in the partition function can be precisely turned off in a modified model. The density matrix renormalization group method is then employed to study these two models comparatively on finite-size systems, which is designed to unveil the consequences of the phase-string component. We find that the hole pairing is present not only in the quasi-long-range superconducting phase but also in the stripe phase of the $t$-$t'$-$J$ model. However, once the phase-string is switched off, both the superconducting and stripe orders together with the underlying hole pairing disappear. The corresponding ground state reduces to a trivial Fermi-liquid-like state with small hole Fermi pockets that is decoupled from the antiferromagnetic spin background. It is in sharp contrast to the original $t$-$t'$-$J$ model where large Fermi surfaces can be restored in the stripe phase found at $t'/t<0$ or the superconducting phase at $t'/t>0$ in the six-leg ladder calculation. Our study clearly demonstrates that the strong correlation effect in doped Mott insulator should be mainly attributed to the long-range quantum entanglement between the spin and charge, which is, non-perturbatively, beyond a simple spin-charge separation under the no double occupancy constraint.
\end{abstract}

\maketitle

\section{\label{sec:ITR} INTRODUCTION}
Understanding the emergence of unconventional superconductivity (SC) and its pairing mechanism is a major task in condensed matter physics~\cite{Keimer_2015,Proust_2019}. 
Since the unconventional SC is usually realized by doping the parent Mott insulators, the Hubbard and effective $t$-$J$ models are commonly taken as the minimal models to study SC in doped Mott insulators~\cite{Keimer_2015,Proust_2019,Anderson_1987,Anderson_2004,Wen_2006,IJMPB2007,Fukuyama2008,Arovas_2022}.
While analytical solutions for two-dimensional correlated systems may be still not well controlled, numerical calculations have found different quantum phases in the doped square-lattice Hubbard and $t$-$J$ models~\cite{White_PRL_1998,White_1999,White2003,Hager2005,Corboz_PRL_2014,Simons_PRX_2015,Ehlers2017,Garnet_Science_2017,EWHuang2017,Ido2018,Jiang_2018,Corboz_2019,QinMingPu_PRX_2020,Mingpu_2022,Xu2022,Martins_2001,Sorella2002,Shih2004,White_2009,Eberlein2014,Kivelson_PRB_2017,HCJ_Science_2019,Chung_PRB_2020,Jiang_PRR_2020,Lu_2022,White_PRB_2022,White_PNAS_2021,Jiang_PRL_2021,Gong_PRL_2021,Jiang_Kivelson_Lee_2023,Wietek_PRL_2022,Wietek2021,Qu2022,Jiang2023,Xu2023}. In particular, for quasi-1D systems some phases have been established, including the spin and charge intertwined charge density wave (CDW) order with relatively weak SC correlation~\cite{White_PRL_1998,White_1999,White2003,Hager2005,Corboz_PRL_2014,Simons_PRX_2015,Ehlers2017,Garnet_Science_2017,EWHuang2017,Ido2018,Jiang_2018,Corboz_2019,QinMingPu_PRX_2020,Mingpu_2022,Xu2022}, as well as the Luther-Emery liquid with coexistent quasi-long-range SC and CDW orders~\cite{Luther_PRL_1974,Balents_PRB_1996} by turning on the next-nearest-neighbor (NNN) hopping $t'$ with both positive and negative $t'/t$, where $t$ is the nearest-neighbor (NN) hopping~\cite{Kivelson_PRB_2017,HCJ_Science_2019,Chung_PRB_2020,Jiang_PRR_2020,Lu_2022}.

Towards wider systems, recent density matrix renormalization group (DMRG) studies on six-leg $t$-$t'$-$J$ model~\cite{White_PRB_2022,White_PNAS_2021,Jiang_PRL_2021,Gong_PRL_2021,Jiang_Kivelson_Lee_2023,Wietek_PRL_2022} find that while the CDW phase persists at $t'/t < 0$ near the optimal doping~\cite{White_PRB_2022,White_PNAS_2021}, it gives way to a robust $d$-wave SC phase with tuning $t'/t>0$ ~\cite{White_PNAS_2021,Jiang_PRL_2021,Gong_PRL_2021,Jiang_Kivelson_Lee_2023}.   
The Fermi surface topology results~\cite{Gong_PRL_2021,White_PNAS_2021} indicate that this SC phase may provide a natural understanding of the cuprates at electron doping, and the $t'/t<0$ side should correspond to the hole doping~\cite{Kim1998,Pavarini2001,Tanaka2004}, thus leaving the emergent SC at hole-doped cuprates still puzzling.
On the other hand, DMRG results suggest that hole binding may also exist in the CDW phase~\cite{White_PNAS_2021}.
These significant numerical results naturally raise several important questions such as what is the origin of hole binding? Whether the holes are indeed paired in the CDW phase, and if they are, what leads to the different SC and CDW phases with tuning $t'/t$?

\begin{figure}[t]
	\includegraphics[width=0.9\linewidth]{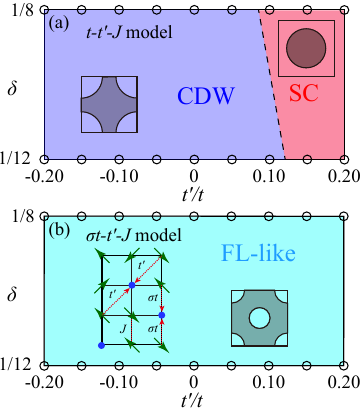}
	\caption{\label{fig:model}Phase diagrams of the $t$-$t'$-$J$ model (a) and $\sigma t$-$t'$-$J$ model (b), determined on the $L_{y}=6$ cylinder. At the range of $-0.2 \leq t^\prime / t \leq 0.2$ and doping level $1/12\leq \delta \leq 1/8$, the CDW and SC phases present in the $t$-$t'$-$J$ model are replaced by a ubiquitous FL-like state in the $\sigma t$-$t'$-$J$ model. The topology of large Fermi surfaces in the CDW and SC phases in (a) are reduced to that of small hole pockets in the FL-like phase in (b), as illustrated by the cartoon plots of the momentum distribution (the darker regions indicate the electron occupancy with higher probability). Here the $\sigma t$-$t'$-$J$ model is distinct from the $t$-$t'$-$J$ model only by a spin-dependent sign in the NN hopping integral $\sigma t$ with the same NNN hopping $t'$ and NN superexchange $J$, as indicated in the left inset of (b) with the arrows (circles) denoting electrons (holes). Following Ref.~\cite{Gong_PRL_2021}, the phase boundary in the $t$-$t'$-$J$ model is determined by examining charge density profile and comparing different correlation functions.}
\end{figure}

The present paper is composed of two combined precise studies of the doped Mott physics governed by the $t$-$t'$-$J$ model. By analytically identifying the sign structure of the model, it is rigorously shown the conventional Fermion statistics is replaced by the phase string or mutual statistics in the $t$-$J$ model ~\cite{Weng.Sheng.1996,Ting.Weng.1997,Zaanen.Wu.2008}, with the NNN hopping $t'$ giving rise to an additional geometric Berry phase. Then by DMRG numerical calculation, it is shown that the phase diagram of the $t$-$t'$-$J$ model at finite doping, including SC and CDW phases, is the consequences of the phase-string sign structure. Namely both the SC and CDW disappear once the phase-string is artificially turned off in DMRG simulation, in which the complex phase diagram reduces to a simple spin-charge separation state irrespective of $t'$: the doped holes form their own small Fermi pockets without pairing on top of the spin background ~\cite{Jiang_Chen_Weng_2020} (see the phase diagrams in Fig.~\ref{fig:model}). It is in sharp contrast to the full $t$-$t'$-$J$ model with the phase-string. Here a \emph{large} Fermi surface is restored, which strongly depends on $t'$. Especially holes are shown to be paired at either $t'/t>0$ or $t'/t<0$ underlying both SC and CDW phases. Our findings demonstrate that an intrinsic quantum entanglement between the spin and charge degrees of freedom via the hidden phase-string, which goes beyond a naive spin-charge separation description with no-double-occupancy constraint, should be crucially important in understanding the doped Mott insulator.

In Sec. II, a precise sign structure in the partition function will be analytically determined for the $t$-$t'$-$J$ model on the square lattice. Such a sign structure replaces the conventional Fermi sign structure in the weak-coupling limit. In contrast to the singular phase-string sign structure for the $t$-$J$ model, it is shown that the NNN hopping $t'$ will lead to an additional Berry phase. On the other hand, in the so-called $\sigma t$-$t'$-$J$ model, the singular phase-string component can be exactly removed in the sign structure with the rest remaining unchanged in the partition function including the Berry phase associated with $t'$. In Sec. III, a comparative DMRG calculation is carried out for both $t$-$t'$-$J$ and $\sigma t$-$t'$-$J$ models on four- and six-leg ladders at finite doping. The ground state properties, including the phase diagrams, pairing energies, Fermi surfaces, CDW profile and correlations, have been systematically explored with and without the phase-string effect. Finally discussion and conclusion are made in Sec. IV.

\section{\label{sec:MH} Model and THEORETICAL FRAMEWORK}
We first discuss the sign structure in the $t$-$t'$-$J$ model. 
The model is defined as $H_{t\text{-}t^\prime \text{-}J}=H_{t\text{-}t^\prime} +H_J$, with both the NN $\langle i j\rangle$ and NNN $\langle\langle i j\rangle\rangle$ hopping terms 
\begin{equation}\label{Ht}
  H_{t-t^{\prime}}\equiv - t \sum_{\langle i j\rangle \sigma}  c_{i, \sigma}^{\dagger} c_{j, \sigma}-t^{\prime} \sum_{\langle\langle i j\rangle\rangle\sigma} c_{i, \sigma}^{\dagger} c_{j, \sigma}+\text {h.c.},
\end{equation}
as well as the NN superexchange term
$H_J = J \sum_{\langle ij \rangle}({\bf S}_i \cdot {\bf S}_j - \frac{1}{4} {n}_i {n}_j)$,
where ${c}^{\dagger}_{i,\sigma}$ and ${c}_{i,\sigma}$ are the creation and annihilation operators for the electron with spin $\sigma/2$ ($\sigma = \pm 1$) at the site $i$, ${\bf S}_{i}$ is the spin-$1/2$ operator, and ${n}_i \equiv \sum_{\sigma} {c}^{\dagger}_{i,\sigma} {c}_{i,\sigma}$ is the electron number operator.
The Hilbert space for each site is constrained by no double occupancy.

The central theme to be established in this work is that the physics of the $t$-$t'$-$J$ model is essentially dictated by the sign structure of the model just like that a FL state is determined by the Fermi sign structure (statistics) in a conventional weakly interacting system. Here the sign structure refers to the sign factor $\tau_C$ in the following partition function for the system at a finite hole doping
\begin{equation}\label{Zex}
  Z_{t\text{-}t^\prime \text{-}J}\equiv \operatorname{Tr} e^{-\beta H_{t\text{-}t^\prime \text{-}J}} =\sum_{C} \tau_{C} W_{t\text{-}t^\prime \text{-}J} [C],
\end{equation}
where $\beta$ is the inverse temperature, $W_{t\text{-}t^\prime \text{-}J} [C]\geqslant 0$ denotes the positive weight for each closed loop $C$ of all spin-hole coordinates on the square lattice, and a quantum sign associated with the path $C$ is characterized by $\tau_{C}=\pm 1$. Here both $W$ and $\tau$ may be generally determined via the stochastic series expansion (cf. Appendix~\ref{appsec:analytic}):
\begin{equation}\label{highZ}
  Z_{t\text{-}t^\prime \text{-}J}=\sum_{n=0}^{\infty} \sum_{\left\{\alpha_{i}\right\}} \frac{\beta^{n}}{n !} \prod_{i=0}^{n-1}\left\langle\alpha_{i}|(-H_{t\text{-}t^\prime \text{-}J})| \alpha_{i+1}\right\rangle,
\end{equation}
where for each $n$, $|\alpha_{n}\rangle=|\alpha_{0}\rangle$ denoting the many-body hole and spin-Ising bases (with $\hat{z}$ as the quantization axis) such that the partition function is characterized by a series of closed loops of step $n$'s, denoted by $C$, which include both hopping and superexchange processes. The total sign collected by $\tau_{C}$ is precisely given by  
\begin{equation}\label{tau}
  \tau_{C}\equiv \tau^0_{C}  \times  (-1)^{N^h_{\downarrow}} 
\end{equation}
with $  \tau^0_{C}\equiv (-1)^{N^h_{\mathrm{ex}}}\times \left[\operatorname{sgn} \left(t^{\prime}\right) \right]^{N^h_{t'}} $, in which $N^h_{\mathrm{ex}}$ denotes the total number of exchanges between the identical holes, i.e., the usual Fermi statistical sign structure of the doped holes like in a semiconductor, and $N^h_{t'}$ counts the total steps of the NNN hopping of the holes in the path $C$, resulting in a geometric Berry phase at $t'<0$.  In Eq.~\eqref{tau}, the NN hopping integral is assumed to be always positive for simplicity, i.e., $t>0$.

\begin{figure*}[t]
\centering
	\includegraphics[scale=0.8]{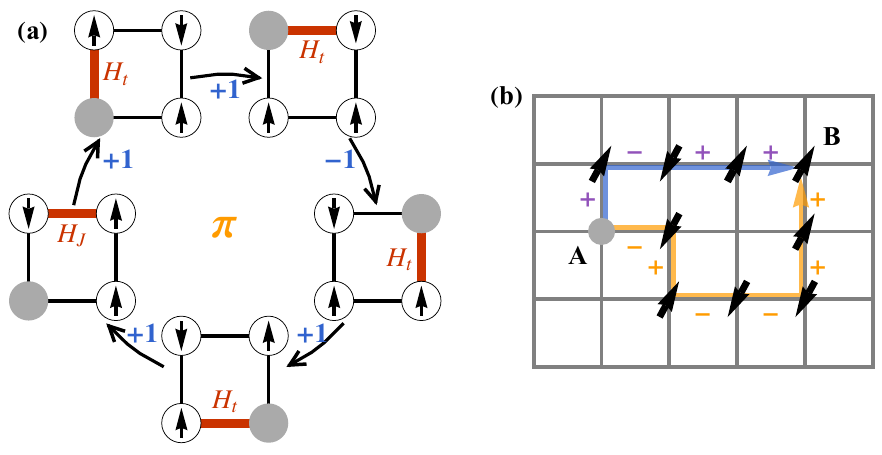}
	\caption{\label{piphase}(a) Illustration of an example of a closed loop for the doped square lattice, with matrix elements of the interaction process ($-H$) labeled by thick red lines. The $\pm1$ markers indicate the sign of the matrix elements. Additionally, the grey circles represent holes. (b) Depiction of the phase-string Berry phase experienced by single holes for different paths.}
\end{figure*}

The sign factor $(-1)^{N^h_{\downarrow}}$ in Eq.~\eqref{tau} is known as the phase-string~\cite{Weng.Sheng.1996,Ting.Weng.1997,Zaanen.Wu.2008}, in which $N^h_{\downarrow}$ denotes the total mutual exchanges between the holes and down-spins at the NN hoppings. 
Thus, a sequence of signs $( \pm 1) \times( \pm 1) \times \cdots \times( \pm 1)$ is picked up by the hole from the spin background, as shown in Fig. \ref{piphase}. Consider a hole moving on a closed path $C$, with the spin displacement induced by its movement reverting to its initial state once the hole returns to its starting point. Consequently, the system reverts to its original state, but with an added phase of $(-1)^{N_{\downarrow}^h}$. This accumulated sign bears a resemblance to a Berry phase. However, unlike the traditional definition of a Berry phase, which requires adiabatic evolution, this phenomenon occurs for any path $C$ as described in Eq.~\eqref{tau}. For instance, as illustrated in Fig.~\ref{piphase}(a), a total phase of $\pi$ is accrued after completing the loop configuration presented. This accumulation of phase-string Berry phase can lead to a considerable obstruction in the coherent propagation of the bare electron (or hole) due to the destructive interference effects among distinct path (shown in Fig. \ref{piphase}(b)), as formulated in the path integral framework. Moreover, such a phenomenon has been previously identified in the $t$-$J$ model at $t'=0$~\cite{Zaanen.Wu.2008} as a form of novel long-range entanglement or mutual statistics between the doped holes and spins, as proposed in earlier studies ~\cite{Ting.Weng.1997,IJMPB2007}.

The phase-string sign structure is thus proven to be generally present in the $t$-$t'$-$J$ model, whereas $t'<0$ merely gives rise to a conventional geometric Berry phase in Eq.~\eqref{tau}. The origin of the phase-string can be traced to the constraint of no double occupancy, i.e., the Mott insulator upon doping, which is protected by the Mott gap. Note that it has nothing to do with the detailed spin states as the result in Eq.~\eqref{tau} holds true at an arbitrary temperature, doping, and sample size. In other words, it must be present to replace the usual Fermi statistics under the no-double-occupancy constraint, whose effect has been omitted in the original construction of the Resonating Valence Bond (RVB) theory~\cite{Anderson_1987} in a doped Mott insulator. Basically the phase-string will imply two important consequences. Besides its novel Berry-phase-like topological effect discussed above, the phase-string also represents a very singular effect at short distances as the NN hopping integral must be modulated by a $\pm$ sign pending on each spin-$S^z$ to be exchanged with a hole. Namely, the dynamic spin background can drastically modify the hole hopping, where a usual perturbative treatment is expected to fail.
So the strong correlation in the $t$-$t'$-$J$ model will be encoded in the phase-string effect.
Interestingly, the pairing mechanism may be straightforwardly understood by observing that a tightly-pairing of two holes can effectively compensate the severe phase-string effect induced by them. Such a scenario has been verified by variational Monte Carlo (VMC) calculations~\cite{Weng.Chen.2018, Weng.Zhao.2022},  and will get further support from the unbiased DMRG results demonstrated below. Notably, this phase-string-induced pairing mechanism stems directly from the strong Coulomb repulsion of the Mott insulator, rather than from spin fluctuations~\cite{Scalapino.Scalapino.2012, Hirsch.Scalapino.1986, Emery.Bal-Monod.1986} or phonons~\cite{zhangchao_eph_2023,eph2_shenzhixun_2021} as intensely studied in previous works.

To explicitly see the effect of the phase-string, we point out that if the hopping term $H_{t-t^{\prime}}$ is changed to 
\begin{equation}\label{sigma t}
  H_{\sigma t-t^{\prime}}\equiv - t \sum_{\langle i j\rangle \sigma} \sigma c_{i, \sigma}^{\dagger} c_{j, \sigma}-t^{\prime} \sum_{\langle\langle i j\rangle\rangle\sigma} c_{i, \sigma}^{\dagger} c_{j, \sigma}+\text {h.c.},
\end{equation}
with the superexchange term $H_J$ unchanged, which is to be called the $\sigma t$-$t'$-$J$ model below, the $\tau_{C}$ in Eq.~\eqref{tau} will reduce to the sole sign structure in the partition function
\begin{equation}\label{Zsigma}
  Z_{\sigma t\text{-}t^\prime \text{-}J} =\sum_{C} \tau^0_{C} W_{t\text{-}t^\prime \text{-}J} [C]~.
\end{equation}
In other words, one can precisely switch off the phase-string in the $\sigma t$-$t'$-$J$ model such that the sign structure becomes a conventional $ \tau^0_{C}$ with the \emph{same} amplitude $W_{t\text{-}t^\prime \text{-}J} [C]$. 
We would like to emphasize that the above sign structures for both models are exact at any finite size, arbitrary doping, and temperature (see the details in Appendix~\ref{appsec:analytic}).

\section{\label{sec:Res} Numerical results}

\subsection{Phase diagrams}

In the following, we shall determine the ground states of both models using DMRG calculations~\cite{White_DMRG_1992} for finite-size cylinder systems, to reveal the important role played by the phase-string component of the sign structure in hole pairing.
We choose $t/J = 3$ to mimic a large Hubbard $U/t = 12$ and tune the hopping ratio in the region of $-0.2 \leq t^\prime / t \leq 0.2$. 
We focus on two doping levels at $\delta = N_h / N = 1/12$ and $1/8$, where $N_h$ is the hole number and $N$ is the total site number.
We study the models on cylinder geometry with the periodic boundary condition along the circumference direction ($y$) and open boundary along the axis direction ($x$), and use $L_y$ and $L_x$ to denote the site numbers along the two directions, respectively.
For the $t$-$t'$-$J$ model with spin SU(2) symmetry, we keep the bond dimensions up to $12000$ SU(2) multiplets~\cite{McCulloch2002} (equivalent to about $36000$ U(1) states). 
For the $\sigma t$-$t'$-$J$ model, we use the U(1) symmetry and keep the bond dimensions up to $15000$ states.
We obtain accurate local measurements and correlation functions on the $L_y = 4, 5, 6$ cylinders with the truncation error near $1\times 10^{-6}$.

\begin{figure*}[t]
	\includegraphics[width=0.75\linewidth]{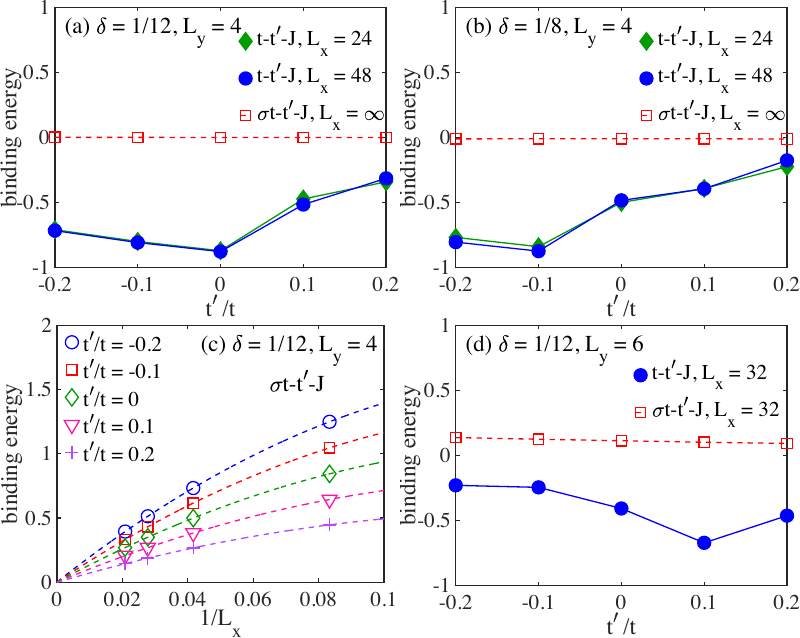}
	\caption{\label{fig:binding}Binding energy in the $t$-$t'$-$J$ and $\sigma t$-$t'$-$J$ models. (a) and (b) show the binding energies on the four-leg cylinders at $\delta=1/12$ and $\delta=1/8$, respectively. The results of the $t$-$t'$-$J$ and $\sigma t$-$t'$-$J$ models are marked with solid and open symbols, respectively. We present two sizes ($L_{x}=24$ and $48$) for the $t$-$t'$-$J$ model to show the good convergence of the binding energy with $L_x$. For the $\sigma t$-$t'$-$J$ model, the binding energies exhibit a pronounced finite-size effect, and we show the results after the size extrapolation to the infinite-$L_{x}$ limit. (c) Extrapolations of binding energies versus system length for the four-leg $\sigma t$-$t'$-$J$ model at $\delta=1/12$, which are fitted by a second-order polynomial function $\mathcal{C} \left(1/L_x \right)=\mathcal{C} \left(0\right)+a/L_x +b/L_x^2$. (d) Binding energies for the two models on the $L_y = 6, L_x = 32$ cylinder at $\delta=1/12$. The results of the $\sigma t$-$t'$-$J$ model are obtained using $12000$ bond dimensions. For the $t$-$t'$-$J$ model, the binding energies are calculated using the energies $E(N_h, S)$ after the extrapolation to the infinite-bond-dimension limit (cf. Fig.~\ref{ground_energy} and Fig.~\ref{ground_energy_5} in Appendix~\ref{appsec:binding}).}
\end{figure*}

The phase diagram of the $t$-$t'$-$J$-$J'$ model on the four-leg system has been established by DMRG in Ref.~\cite{Jiang_PRR_2020}, which should be qualitatively consistent with the $t$-$t'$-$J$ model because $J'$ is much smaller.
Here, we illustrate the phase diagrams obtained on the six-leg cylinder in Fig.~\ref{fig:model}. 
With the phase-string, the $t$-$t'$-$J$ model shows the CDW and SC phases as a function of $t'/t$ [Fig.~\ref{fig:model}(a)], both of which have large Fermi surfaces.
By contrast, by switching off the phase-string, the $\sigma t$-$t'$-$J$ model exhibits a FL-like phase with small Fermi pockets in the whole studied range of $t'/t$ [Fig.~\ref{fig:model}(b)], which is dictated by the Fermi signs in $ \tau^0_{C}$.

\subsection{Binding energy}
An important result we find in this work is that in the $t$-$t'$-$J$ model the holes form pairs not only in the SC phase, but also in the CDW phase.
Following the usual definition, we calculate the binding energy $E_b$ as
\begin{equation} \label{eq:binding}
E_b \equiv E(N_h, 0) + E(N_h-2, 0) - 2 E(N_h-1, 1/2),
\end{equation}
where $E(N_h, S)$ denotes the lowest-energy in the sector with hole number $N_h$ and total spin quantum number $S$ (total spin-$z$ component $S$) for the $t$-$t'$-$J$ ($\sigma t$-$t'$-$J$) model.
The negative $E_b$ characterizes the existence of hole binding.
On the four-leg cylinder, our large bond dimensions can guarantee a very good convergence of binding energy.
In the $t$-$t'$-$J$ model, although the system can exhibit Luther-Emery liquid and filled stripe phases with tuning parameters~\cite{Jiang_PRR_2020}, the binding energies, which converge quickly with system length (see the consistent $L_x = 24$ and $48$ results in Figs.~\ref{fig:binding}(a) and \ref{fig:binding}(b)), are always negative.
On the other hand, for the $\sigma t$-$t'$-$J$ model, the binding energies are positive and strongly size dependent.
Importantly, the positive binding energies of the $\sigma t$-$t'$-$J$ model are clearly extrapolated to zero for $L_x \to \infty$ [see Fig.~\ref{fig:binding}(c), and Fig.~\ref{fig:BE_8_extrap}(a) in Appendix~\ref{appsec:binding} at $\delta = 1/8$].

To verify the distinct pairing properties of the two models, we also calculate the binding energy of the wider systems.
On the five-leg $t$-$t'$-$J$ model (cf. Figs.~\ref{fig:BE_8_extrap}(b) and \ref{fig:BE_8_extrap}(c) in Appendix~\ref{appsec:binding}), the binding energies are also negative and have very small finite-size effects in our studied system length, which are consistent with the quasi-long-range SC order (see the correlation function results in Appendix~\ref{appsec:fiveleg}). 
Given the negligible $L_x$ dependence of binding energy in the $t$-$t'$-$J$ model, we study the six-leg system only at $L_x = 32$ due to the much larger computational cost, which show qualitatively the same conclusion as presented in Fig.~\ref{fig:binding}(d) for $\delta = 1/12$ and Fig.~\ref{fig:BE_8_extrap}(d) for $\delta = 1/8$.
As a complementary check, we have also calculated the binding energy of the three-leg cylinder [see Figs.~\ref{fig:BE_8_extrap}(e) and \ref{fig:BE_8_extrap}(f)].
The systems also have small negative binding energy that show a weak size dependence.
Therefore, our results strongly suggest the hole binding as a universal feature on both even- and odd-leg $t$-$t'$-$J$ model.
For computing the binding energy in the $t$-$t'$-$J$ model, the energies $E(N_h, S)$ in Eq.~\eqref{eq:binding} have all been extrapolated to the infinite-bond-dimension limit (see Fig.~\ref{ground_energy} and Fig.~\ref{ground_energy_5} in Appendix~\ref{appsec:binding}).
For the $\sigma t$-$t'$-$J$ model, the obtained binding energies, which are almost independent of the kept bond dimension, have small positive values around $0.1$ on the $L_y = 6, L_x = 32$ cylinder, much smaller than the values for the same $L_x$ on the four-leg cylinder and strongly suggesting the vanished binding energy in the infinite-$L_x$ limit.
Notice that the positive binding energy also indicates the absent hole pairing in the finite-$L_x$ systems.

Since the sign structures for both models are exact at any finite size, our DMRG results at $L_y = 3, 4, 5, 6$ provide a strong evidence to support the crucial role of the phase-string component for hole pairing.
Remarkably, while the pairing correlation decays exponentially in the CDW phase of the six-leg $t$-$t'$-$J$ model [Fig.~\ref{fig:correlation}(c)], the hole pairing still exists, which is quite consistent with the characteristic of a pseudogap phase~\cite{IJMPB2007}. 
Next, we will further discuss this point from the perspective of correlation functions. 
Our results show that the hole pairing is indeed diminished no matter in what phase if one switches off the phase-string precisely.

At the end of this subsection, we discuss the width dependence of binding energy in the $t$-$t'$-$J$ model (see Fig.~\ref{fig:BEE}). While our size-limited data may not determine the existence of hole pairing in the thermodynamic limit, the positive signal is that the binding generally becomes stronger from five- to six-leg system.

\subsection{Fermi surface}
\begin{figure*}[t]
	\includegraphics[width=0.65\linewidth]{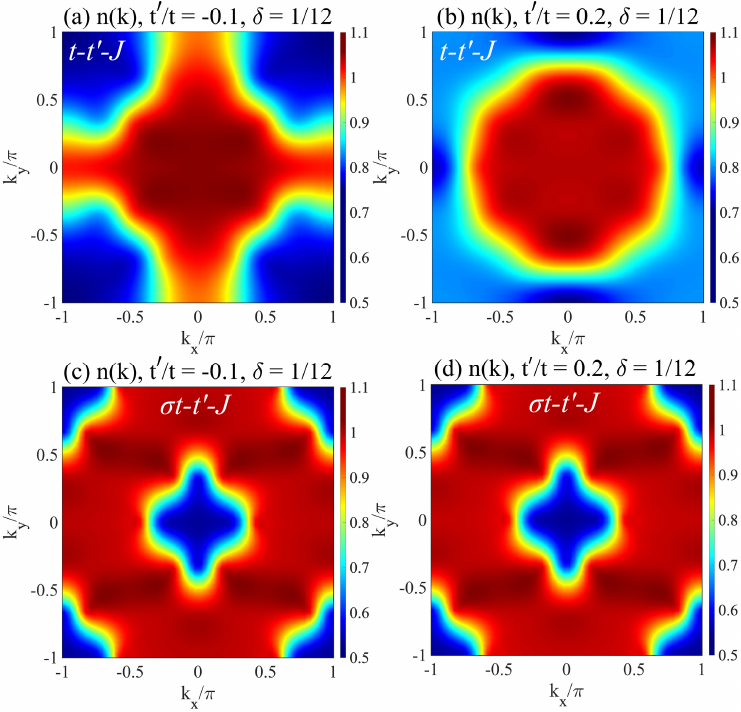}
	\caption{\label{fig:nk}Electron distribution in the momentum space $n(\bf k)$ on six-leg cylinder with $\delta = 1/12$. (a) and (b) show the results of the $t$-$t'$-$J$ model with a large Fermi surface. The Fermi surface topology is different in the CDW and SC phases. (c) and (d) show the results of $\sigma t$-$t'$-$J$ model. The electrons of spin-up and spin-down in $n(\bf k)$ are displaced by $(\pi,\pi)$.}
\end{figure*}

In Fig.~\ref{fig:nk}, we demonstrate the electron momentum distributions $n\left(\mathbf{k}\right)=\frac{1}{N}\sum_{i,j,\sigma} \langle {\hat{c} }_{i,\sigma}^{\dagger } {\hat{c} }_{j,\sigma} \rangle e^{i\mathbf{k}\cdot \left({\mathbf{r}}_i -{\mathbf{r}}_j \right)}$. 
The $t$-$t'$-$J$ model always exhibits a large Fermi surface with hole binding as given above, but the topology of Fermi surface has a strong $t'/t$ dependence.
By contrast, the $\sigma t$-$t'$-$J$ model shows two small Fermi pockets at $\mathbf{k}=\left(\pi ,\pi \right)$ and $\left(0 ,0 \right)$, which have a weak $t'/t$ dependence and are contributed from the spin-up and spin-down propagators, respectively (see Appendix~\ref{appsec:nk} for the results of other parameters). 
For both spin components, the quasi-long-range single-particle correlations $G_{\sigma}(r) \equiv \langle c^\dagger_{(x,y),\sigma}  c_{(x+r,y),\sigma}\rangle \sim r^{-K_{\mathrm{G}}}$ with $K_{\mathrm{G}} \approx 1$ [Fig.~\ref{fig:correlation}(d)] resemble a FL-like state, which smoothly persists to a finite $t^\prime/t$ (cf. Table \ref{Table I}).
Such a non-pairing state, which has been reported at $t'/t=0$ in the two-leg and four-leg systems~\cite{Jiang_Chen_Weng_2020}, appears to be insensitive to $t'/t$ and stable on larger system size.

\subsection{CDW profile and correlation functions}
\begin{figure*}[t]
	\includegraphics[width=0.75\linewidth]{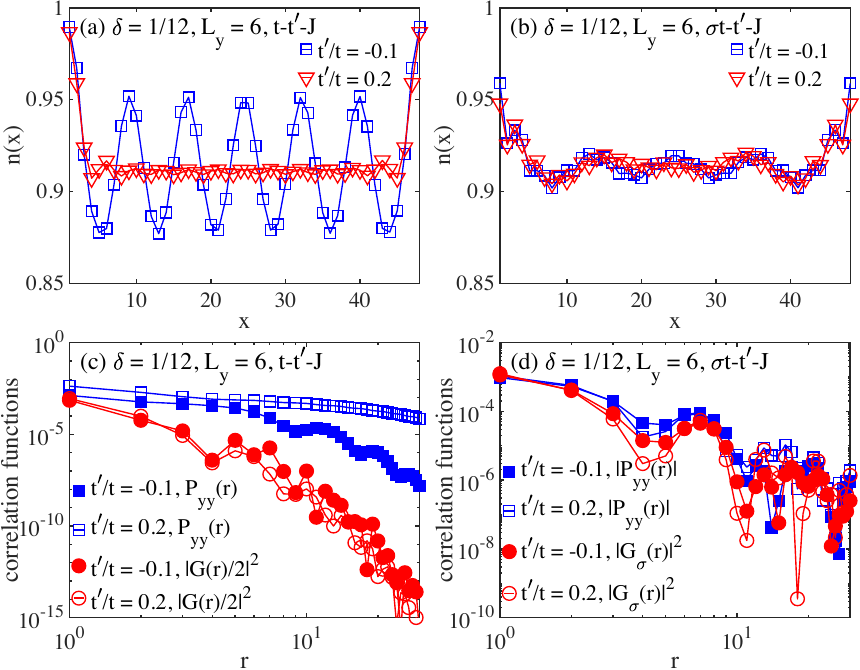}
	\caption{\label{fig:correlation}Charge density profile and correlation functions on the six-leg cylinder with $L_x = 48$ and $\delta=1/12$. (a) and (b) show the charge density profiles $n(x)$ for the $t$-$t'$-$J$ and $\sigma t$-$t'$-$J$ models, respectively. (c) and (d) are the double logarithmic plots of the pairing correlation $P_{yy}(r)$ and single-particle Green's function $G(r)$ for the $t$-$t'$-$J$ and $\sigma t$-$t'$-$J$ models. Note that we only show the $\sigma$ component of the single-particle Green's function in the $\sigma t$-$t'$-$J$ model, due to the absence of spin SU(2) symmetry~\cite{Jiang_Chen_Weng_2020}. The results of more correlation functions and the fitted decay exponents are summarized in Table~\ref{Table I}.}
\end{figure*}

Next, we characterize the two systems through charge density profile and correlation functions.
We measure the average charge density distribution in the column $x$ as $n(x) = \sum_{y=1}^{L_y} \langle \hat{n}(x,y)\rangle / L_y$.
For the $t$-$t'$-$J$ model on the six-leg cylinder, one can see a clear charge density oscillation with a period of $4 / (L_y \delta)$ in the CDW phase and uniform charge distribution in the SC phase [Fig.~\ref{fig:correlation}(a)].
On the other hand, the charge density distribution has no apparent dependence on $t'/t$ in the $\sigma t$-$t'$-$J$ model, and the charge oscillation is quite weak and without a well-defined periodicity by switching off the phase-string [Fig.~\ref{fig:correlation}(b)].

\begin{table*}
   \caption{Summary of the quantum phases for the $t$-$t'$-$J$ and $\sigma t$-$t'$-$J$ model on the six-leg cylinder with doping ratio $\delta=1/12$. The decay exponents of correlation functions are presented at $t'/t = -0.1, 0, 0.2$. The pairing correlation $P_{yy}(r)$ and single-particle Green's function $G(r)$ have been defined in the text. Here we also show the decay exponents of the density correlation function $D(r) = \langle {\hat{n} }(x,y) {\hat{n} }(x+r,y) \rangle -\langle {\hat{n} }(x,y) \rangle \langle {\hat{n} }(x+r,y) \rangle$ and spin correlation function $F(r) = \langle {\mathbf{S}}(x,y) \cdot {\mathbf{S}}(x+r,y) \rangle$. The correlation length of exponential fitting is denoted as $\xi$, and the power exponent of algebraic fitting is denoted by $K$. The fitted correlation functions for the $t$-$t'$-$J$ and $\sigma t$-$t'$-$J$ models are obtained by keeping $12000$ SU(2) multiplets and $15000$ U(1) states, respectively. For the FL-like phase in the $\sigma t$-$t'$-$J$ model, the pairing correlation $P_{yy}(r)$ will behave as a product of two Green's functions and thus also follows an algebraic decay, as shown in Fig.~\ref{fig:correlation}(d). Nonetheless, this algebraic decay does not characterize a quasi-long-range SC order. Therefore, we do not fit the power exponent $K_{\rm sc}$ in the FL-like phase.}
     \begin{ruledtabular}\label{Table I}
        \begin{tabular}{c l c c c c c}
         Models & Parameters & Phase & $P_{yy}(r)$ & $D(r)$ & $G(r)$ & $F(r)$ \\
         \hline
                                              & $t^{\prime}/t=-0.1$ & CDW & $\xi_{\mathrm{sc}} \approx 3.02$ & $\xi_{\mathrm{c}} \approx 4.72$ & $\xi_{\mathrm{G}} \approx 2.73$ & $\xi_{\mathrm{s}} \approx 5.91$ \\
 $t$-$t^{\prime}$-$J$                                & $t^{\prime}/t=0$ & CDW & $\xi_{\mathrm{sc}} \approx 2.95$ & $\xi_{\mathrm{c}} \approx 4.96$ & $\xi_{\mathrm{G}} \approx 2.03$ & $\xi_{\mathrm{s}} \approx 6.13$\\
                                              & $t^{\prime}/t=0.2$ & SC + CDW & $K_{\mathrm{sc}} \approx 0.55$ & $K_{\mathrm{c}} \approx 1.56$ & $\xi_{\mathrm{G}} \approx 1.97$ & $\xi_{\mathrm{s}} \approx 3.32$ \\
          \hline                                    
                                              & $t^{\prime}/t=-0.1$ & FL-like & --- & $K_{\mathrm{c}} \approx 1.93$ & $K_{\mathrm{G}} \approx 0.96$ & $K_{\mathrm{s}} \approx 1.83$ \\
$\sigma t$-$t^{\prime}$-$J$        & $t^{\prime}/t=0$ & FL-like & --- & $K_{\mathrm{c}} \approx 1.55$ & $K_{\mathrm{G}} \approx 0.81$ & $K_{\mathrm{s}} \approx 1.69$\\
                                              & $t^{\prime}/t=0.2$ & FL-like & --- & $K_{\mathrm{c}} \approx 1.82$ & $K_{\mathrm{G}} \approx 0.94$ & $K_{\mathrm{s}} \approx 1.95$\\
         \end{tabular}
      \end{ruledtabular}
\end{table*}

We further compare different correlation functions.
We examine the spin-singlet pairing correlation between the vertical bonds $P_{y,y}(r) \equiv \langle \hat \Delta^\dagger_{y}(x,y) \hat \Delta_{y}(x+r,y)\rangle$, 
where the pairing operator is defined for two NN sites $(x,y)$ and $(x,y+1)$, i.e. $\hat \Delta_{y}(x,y) = (c_{(x,y),\uparrow}c_{(x,y+1),\downarrow} - c_{(x,y),\downarrow}c_{(x,y+1),\uparrow})/\sqrt{2}$.
In the $t$-$t'$-$J$ model [Fig.~\ref{fig:correlation}(c)], the pairing correlation decays algebraically with strong magnitudes in the SC phase but is suppressed to decay exponentially in the CDW phase.
Remarkably, the weakened pairing correlation in the CDW phase is still much stronger than two single-particle correlator $G^{2}(r)$, which could be consistent with a pseudogap phase with hole binding but lacking phase coherence due to the strong CDW. 
Such pseudogap-like behaviors have also been observed in the triangular-lattice $t$-$J$ model~\cite{Huang_2022}, which also sits near a SC phase and may be common in doped Mott insulators.


For the $\sigma t$-$t'$-$J$ model, we find that all the correlations exhibit a nice algebraic decay (see Fig.~\ref{fig:correlation}(d) and Table~\ref{Table I}), and the correlations behave smoothly and consistently as a function of $t'/t$ by switching off the phase-string.
The pairing correlation behaving as $G^2(r)$ [Fig.~\ref{fig:correlation}(d)] agrees with the prediction of a FL-like state and confirms no quasi-long-range SC order. 
Our results indicate no phase transition in the $\sigma t$-$t'$-$J$ model and only a FL-like phase exists.
In particular, the quasi-long-ranged Green's function with $K_{\mathrm{G}} \approx 1.0$ is consistent with the description of the Landau Fermi liquid theory.

\section{Discussion and conclusion}

By using DMRG calculation, we have unveiled that the hole pairs constitute the basic building blocks not only in the SC but also in the CDW phase of the $t$-$t^\prime$-$J$ model. 
We have also identified the precise sign structure of the model, which is composed of the Fermi statistics between the doped holes, a geometric Berry phase depending on the sign of $t^\prime$, and the phase-string mutual statistics between charge and spin degrees of freedom. 
In particular, a mere geometric Berry phase at $t^\prime<0$, with the same amplitude $W_{t\text{-}t^\prime \text{-}J} [C]$ [cf. Eqs.~\eqref{Zex} and \eqref{tau}], may stabilize the stripe over SC order for six-leg cylinders. 
By turning off the phase-string, the hole pairing gets diminished to result in a FL-like phase with no more SC and stripe orders, and the corresponding Fermi surface also drastically reconstructs to become small pockets. This FL-like phase is no longer sensitive to the sign of $t^\prime$. 
The phase-string brings in a strong correlation effect that is responsible for not only the hole pairing~\cite{Weng.Zhao.2022, Weng.Chen.2018} but also restoring a \emph{large} Fermi surface via ``momentum shifting''~\cite{PhysRevB.98.035129, Zhao2022, Zhang.Weng_2022}. 
Namely the one-to-one correspondence principle of the Landau paradigm, which works for weak interaction in the conventional FL/BCS description, is generally violated here. 

In conclusion, as a typical doped Mott insulator, we have identified the phase-string sign structure in the $t$-$t^\prime$-$J$ model and demonstrated that it plays an indispensable role in shaping the novel properties in the model including the pairing mechanism, superconductivity and charge order. Without it, as shown by DMRG in the $\sigma t$-$t^\prime$-$J$ model, the doped holes are essentially decoupled from the spin background to form a Fermi pocket state like in a doped semiconductor. Namely the model would be reduced to a weakly-correlated Fermi gas state for dopants once the phase string is turned off, even though the no-double-occupancy constraint is still present. Therefore, the strongly correlated nature of the $t$-$t^\prime$-$J$ model is represented by the phase-string effect to give rise to superconductivity and pseudogap-type behavior (i.e., the preformed pairing). The present study clearly illustrates that the Mott physics is intrinsically a long-range spin-charge entanglement problem, which goes beyond a simple spin-charge separation from the RVB picture as originally envisaged \cite{Anderson_1987,Wen_2006}.

\begin{acknowledgments}
We acknowledge stimulating discussions with Zheng Zhu and Hong-Chen Jiang. X.~L. and S.~S.~G. were supported by the National Natural Science Foundation of China (No. 12274014), the Special Project in Key Areas for Universities in Guangdong Province (No. 2023ZDZX3054), and the Dongguan Key Laboratory of Artificial Intelligence Design for Advanced Materials (DKL-AIDAM). J.~X.~Z. and Z.~Y.~W. were supported by MOST of China (Grant No. 2021YFA1402101) and NSF of China (Grant No. 12347107).  D.~N.~S. was supported by the US National Science Foundation Grant  No. PHY-2216774.
\end{acknowledgments}

\appendix

\onecolumngrid
\section{The exact sign structure of the $t$-$t^\prime$-$J$ model}
\label{appsec:analytic}

\subsection{Exact sign structure on the square lattice}

In this section, we give a rigorous proof of the partition function in Eq. (\ref{Zex}) and the sign structure given in Eq. (\ref{tau}). We shall start with the slave-fermion formalism, in which the electron operator is defined as $c_{i\sigma}=f_{i}^{\dagger} b_{i\sigma}$, with $f_{i}^{\dagger}$ denoting the fermionic holon operator and $b_{i\sigma}$ denoting the bosonic spinon operator, which satisfies the constraint $f_{i}^{\dagger} f_{i}+\sum_{\sigma} b_{i\sigma}^{\dagger} b_{i \sigma}=1$. To clarify the sign structure of this model, we explicitly incorporate the Marshall sign into the $S_z$-spin representation by implementing the substitution
\begin{equation}\label{Marshall}
b_{i \sigma}\rightarrow (-\sigma)^i b_{i \sigma}
\end{equation}
such that
\begin{equation}\label{frac}
c_{i \sigma}=(-\sigma)^i f_i^{\dagger} b_{i \sigma}.
\end{equation}
Then, the $t$-$t^\prime$-$J$ model can be expressed under this transformation as follows:
\begin{equation}
  H_{t-t^{\prime}-J}=-t\left(P_{o \uparrow}-P_{o \downarrow}\right)-t^{\prime} T_{o}-\frac{J}{4}\left(Q+P_{\uparrow \downarrow}\right),
\end{equation}
where
\begin{eqnarray}
  P_{o \sigma}&=&\sum_{\langle ij\rangle} b_{i \sigma}^{\dagger} b_{j \sigma} f_{j}^{\dagger} f_{i}+\text {h.c.}\\
  T_{o}&=&\sum_{\langle\langle ij\rangle\rangle\sigma}  b_{i \sigma}^{\dagger} b_{j \sigma} f_{j}^{\dagger} f_{i}+\text {h.c.}\\
  P_{\uparrow \downarrow}&=&\sum_{\langle i j\rangle} b_{i \uparrow}^{\dagger} b_{j \downarrow}^{\dagger} b_{i \downarrow} b_{j \uparrow}+\text {h.c.} \\
  Q&=&\sum_{\langle ij\rangle}\left(n_{i \uparrow} n_{j \downarrow}+n_{i \downarrow} n_{j \uparrow}\right).
\end{eqnarray}
Here $P_{o \sigma}$ ($T_{o}$) denotes the hole-spin nearest-neighbor (NNN) exchange operator, $P_{\uparrow \downarrow}$ the nearest-neighbor (NN) spin superexchange operator, and $Q$ describes a potential term between NN spins. By making the high-temperature series expansion [cf. Eq. (\ref{highZ})] of the partition function up to all orders \cite{Zaanen.Wu.2008}
\begin{eqnarray}
  Z_{t\text{-}t^\prime \text{-}J}&=&\operatorname{Tr} e^{-\beta H_{t-t^{\prime}-J}}=\operatorname{Tr} \sum_{n=0}^{\infty} \frac{\beta^{n}}{n !}\left(-H_{t-t^{\prime}-J}\right)^{n}\notag\\
  &=&\sum_{n=0}^{\infty}  \frac{(J \beta / 4)^{n}}{n !} \operatorname{Tr}\left[\sum \ldots\left(\frac{4 t}{J} P_{o \uparrow}\right) \ldots P_{\uparrow \downarrow} \ldots\left(\frac{-4 t}{J} P_{o \downarrow}\right) \ldots\left(\frac{4 t^{\prime}}{J} T_{o}\right) \ldots Q \ldots\right]_{n}\notag\\
  &=&\sum_{n=0}^{\infty}(-1)^{N_{ \downarrow}^h}\left(\operatorname{sgn} t^{\prime}\right)^{N_{t^{\prime}}^h} \frac{(J \beta / 4)^{n}}{n !} \operatorname{Tr}\left[\sum \ldots\left(\frac{4 t}{J} P_{o \uparrow}\right) \ldots P_{\uparrow \downarrow} \ldots\left(\frac{4 t}{J} P_{o \downarrow}\right) \ldots\left(\frac{4 \left| t^{\prime}\right|}{J} T_{o} \right) \ldots Q \ldots\right]_{n}
\end{eqnarray}
where the NN hopping integral is assumed to always be positive for simplicity, that is $t>0$. 
The notation $[\sum \ldots]_n$ indicates the summation over all $n$-block production, and because of the trace, the initial and final hole and spin configurations should be the same such that all contributions to $Z_{t\text{-}t^\prime \text{-}J}$ can be characterized by closed loops of holes and spins. Here $N_{ \downarrow}^h$ denotes the number of NN exchanges between down-spins and holes, as well as $N_{t^\prime}^h$ denotes the number of NNN exchanges between spins and holes, regardless of spin direction. Inserting complete Ising basis with holes
\begin{equation}
  \sum_{\phi\left\{l_{h}\right\}}\left|\phi ;\left\{l_{h}\right\}\right\rangle\left\langle\phi ;\left\{l_{h}\right\}\right|=1
\end{equation}
between the operator inside the trace with $\phi$ specifying the spin configuration and $\left\{l_{h}\right\}$ denoting the positions of holes. Then, all the elements inside the trace are positive and the partition function can arrive at a compact expression:
\begin{equation}
  Z_{t\text{-}t^\prime \text{-}J}=\sum_{C} \tau_{C}W_{t\text{-}t^\prime \text{-}J} [C],
\end{equation}
where all the sign information is captured by 
\begin{equation}\label{tautotal}
  \tau_{C}\equiv \tau_C^0 \times(-1)^{N_{\downarrow}^h},
\end{equation}
with
\begin{equation}\label{tau0}
  \tau_C^0 \equiv(-1)^{N_{\mathrm{ex}}^h} \times\left[\operatorname{sgn}\left(t^{\prime}\right)\right]^{N_{t^{\prime}}^h},
\end{equation}
which is consistent with Eq.~\eqref{tau} in the main text. Here, $N_{\mathrm{ex}}^h$ denotes the number of exchanges between holes due to fermionic statistics of holon $f$. Such a sign structure is precisely described at arbitrary doping, temperature, and finite-size for $t\text{-}t^\prime \text{-}J$ model, which has been previously identified at $t^\prime=0$ in Ref. \cite{Zaanen.Wu.2008}. In addition, the non-negative weight $W[C]$ for closed loop $C$ is given by:
\begin{equation}
  W_{t\text{-}t^\prime \text{-}J} [C]=\left(\frac{4 t}{J}\right)^{M_{t}[C]}\left(\frac{4 |t^{\prime}|}{J}\right)^{M_{t^{\prime}}[C]} \sum_{n} \frac{(J \beta / 4)^{n}}{n !} \delta_{n, M_{t}+M_{t^\prime}+M_{\uparrow \downarrow}+M_Q} \geq 0,
\end{equation}
in which $M_t$ and $M_{t^\prime}$ represent the total steps of the hole NN and NNN “hoppings” along the closed loops for a given path $C$ with length $n$, respectively. Also, $M_{\uparrow \downarrow}$ represents the steps of NN spin exchange process,  while $M_Q$ represents the total number of down spins interacting with up spins via the $S_z$ components of the superexchange term.

In summary, the sign structure of the $t$-$t^\prime$-$J$ model comprises not only the effects of NNN hopping and the traditional fermionic statistics encoded in $\tau_C^0$, but also a significant component of $(-1)^{N_\downarrow^h}$ originating from the NN hole hopping process. This component is depicted in Fig.~\ref{piphase}(a) by blue $\pm 1$ on the arrows. From the viewpoint of the original representation, i.e., before the Marshall sign transformation in Eq. \eqref{frac}, each spin flip results in a negative sign under the Ising basis, since $\langle\downarrow_i \uparrow_j|J S_i^{-} S_j^{+}| \uparrow_i \downarrow_j\rangle >0$. Therefore, in the presence of hole hopping, an odd number of spin flips can occur in the closed loop of a bipartite lattice, as illustrated in Fig. \ref{piphase}(a) by black $\pm 1$ on the arrows.

Furthermore, by introducing the $\sigma  t\text{-}t^\prime \text{-}J$ model, in which the original kinetic energy term Eq.~\eqref{Ht} is replaced by:
\begin{equation}
  H_{\sigma t-t^{\prime}}=-\sigma t \sum_{\langle i j\rangle} c_{i, \sigma}^{\dagger} c_{j, \sigma}-t^{\prime} \sum_{\langle\langle i j\rangle\rangle} c_{i, \sigma}^{\dagger} c_{j, \sigma}+\text {h.c.},
\end{equation}
where an extra spin dependent sign $\sigma$ is inserted into the NN hopping term that cancels the ``$-$'' sign in front of the $P_{o \downarrow}$. Consequently, $\sigma  t\text{-}t^\prime \text{-}J$ model under the representation of Eq.~\eqref{frac} can be rewritten as:
\begin{equation}
  H_{t-t^{\prime}-J}=-t\left(P_{o \uparrow}+P_{o \downarrow}\right)-t^{\prime} T_{o}-\frac{J}{4}\left(Q+P_{\uparrow \downarrow}\right),
\end{equation}
with the partition function under the high-temperature series expansion:
\begin{eqnarray}
  Z_{\sigma t\text{-}t^\prime \text{-}J} = \sum_{n=0}^{\infty}\left(\operatorname{sgn} t^{\prime}\right)^{N_{t^{\prime}}^h} \frac{(J \beta / 4)^{n}}{n !} \operatorname{Tr}\left[\sum \ldots\left(\frac{4 t}{J} P_{o \uparrow}\right) \ldots P_{\uparrow \downarrow} \ldots\left(\frac{4 t}{J} P_{o \downarrow}\right) \ldots\left(\frac{4 \left| t^{\prime}\right|}{J} T_{o} \right) \ldots Q \ldots\right]_{n}.
\end{eqnarray}
Consequently, the sign structure for $\sigma t\text{-}t^\prime \text{-}J$ model is given by:
\begin{equation}
  \tau_{C}^{\sigma t\text{-}t^\prime \text{-}J}= \tau_{C}^0,
\end{equation}
where $\tau_{C}^0$ is given by Eq.~\eqref{tau0} and the positive weight $W[C]$ for each path remains unchanged, leading to Eq.~\eqref{Zsigma}.

\subsection{Extension of exact sign structure to non-bipartite lattices }

We have established a rigorous proof of the exact sign structure for the $t$-$t^\prime$-$J$ model. In this subsection, we provide an alternative derivation of the sign structure and demonstrate its applicability to non-bipartite lattices, such as triangular lattices or systems with the NNN superexchange interactions. 

\begin{figure}[t]
	\includegraphics[scale=1.0]{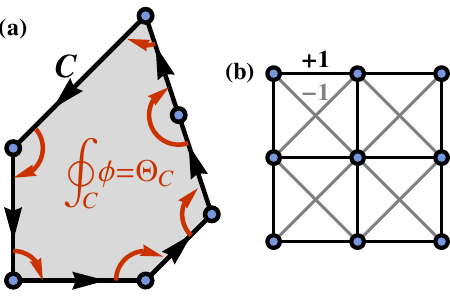}
	\caption{\label{fig_gauge}(a) Illustration the geometric phase $\phi_{ij}$, where the red arrow indicates the non-zero contribution of the points on the selected closed loop (black line), with the total inner angle sum denoted as $\Theta_C$. (b) In the $t$-$t^\prime$-$J$ model, one possible gauge choice for $\phi_{ij}$ is shown here, with black bonds representing nearest-neighbor links having $e^{i\phi_{ij}}=+1$ and gray bonds representing next-nearest-neighbor links having $e^{i\phi_{ij}}=-1$.}
\end{figure}

Here we use the extended $t$-$J$ model on an \emph{arbitrary} lattice as an example, with the Hamiltonian given by
\begin{equation}\label{extended}
    H = - T\sum_{ij,\sigma} c_{i\sigma}^\dagger c_{j\sigma}+h.c.+ K \sum_{ij} \left( \mathbf{S}_i \cdot \mathbf{S}_j-\frac{1}{4} n_i n_j \right).
\end{equation}
Unlike the main text, where the sum indexes $\langle ij \rangle$ represent only the NN links, here the sum includes all allowed links that can connect any two sites. To begin, instead of applying the Marshall basis transformation as described in Eq.~\eqref{Marshall}, which relies on the $A$-$B$ sublattice division, we propose a redefinition of the up-spinon operator as follows: 
\begin{equation}\label{tran}
b_{i \uparrow} \rightarrow b_{i \uparrow} e^{-i \Phi_{i}}
\end{equation}
while keeping the down-spinon operator unchanged, such that
\begin{eqnarray}
    c_{i \uparrow}&=&f_i^{\dagger} b_{i \uparrow} e^{-i \Phi_{i}} \notag \\
    c_{i \downarrow}&=&f_i^{\dagger} b_{i \downarrow},
\end{eqnarray}
where
\begin{equation}
\Phi_{i} \equiv \sum_{l \neq i} \theta_{i}(l)=\sum_{l \neq i} \operatorname{Im} \ln \left(z_{i}-z_{l}\right),
\end{equation}
with $z_i$ as the complex coordinate at site $i$. Hence, by simply using the relations $\theta_{i}(j)-\theta_{j}(i)=\pm \pi$, the extended $t$-$J$ model in Eq.~\eqref{extended} become:
\begin{equation}
  H_{t\text{-}t^{\prime}}=-\sum_{i j, \sigma} \left(\sigma T b_{i \sigma}^{\dagger} b_{j \sigma} f_{j}^{\dagger} f_{i} e^{\frac{i(\sigma+1)}{2} \phi_{ij}}+h.c.\right) -\frac{K}{4} \sum_{\langle ij\rangle \sigma}\left(n_{i \sigma} n_{j-\sigma}+b_{i \sigma}^{\dagger} b_{j-\sigma}^{\dagger} b_{i-\sigma} b_{j \sigma} e^{i \sigma \phi_{i j}}\right)\label{gauge},
\end{equation}
where 
\begin{equation}
   \phi_{ij}=\sum_{l \neq i j}\left[\theta_{i}(l)-\theta_{j}(l)\right]
\end{equation}
acting like a gauge potential with a gauge invariant strength given by $\sum_{C} \phi_{ij}=\Theta_C$ for a closed loop $C$ on the lattice. The symbol $\Theta_C$ denotes the interior angle sum of the closed-loop $C$ depicted in Fig.~\ref{fig_gauge}(a). It is evident that any point that is situated outside or inside the loop contributes $0$ or $\pm 2\pi$ to $\Theta_C$, respectively. Only the sites $l$ on the closed loop $C$ contribute nontrivial values to the interior angle, such that the sum of interior angles over all the points $l \in C$ corresponds to the total contribution, as illustrated in Fig.~\ref{fig_gauge}(a). Importantly, the phase $\phi_{ij}$ presented here is universally applicable to arbitrary lattices, irrespective of whether they are bipartite or not. Notably, for a bipartite system such as the $t$-$J$ model on a square lattice with $t^\prime=0$, $\Theta_C$ is equal to $2\pi \mathbb{Z}$ for any closed loop, and the phase $\phi_{ij}$ can be completely gauged away.

As the result, in Eq.~\eqref{gauge}, the hidden sign structure of extended $t$-$J$ model on an arbitrary lattice is explicitly decomposed into the geometry phase $\phi_{ij}$ and the extra $\sigma$-sign in the hopping term, which is commonly referred to as the ``phase string effect''. To derive the sign structure Eq.~\eqref{tautotal} of $t$-$t^\prime$-$J$ model on a square lattice, a proper gauge can be selected, as shown in Fig.~\ref{fig_gauge}(b), where NN links with $e^{i\phi_{ij}}=+1$ and NNN links with $e^{i\phi_{ij}}=-1$, to combine these two components of frustration, yielding Eq.~\eqref{tautotal}.

\section{More DMRG data for binding energy}
\label{appsec:binding}

\begin{figure*}[t]
	\includegraphics[width=0.9\linewidth]{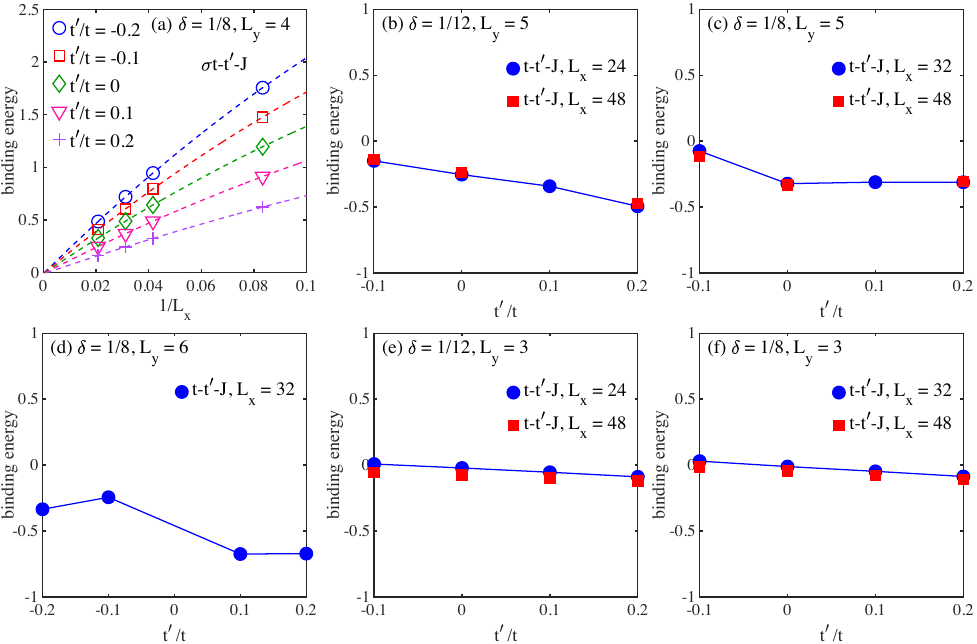}
    \caption{\label{fig:BE_8_extrap}Binding energy in the $t$-$t'$-$J$ and $\sigma t$-$t'$-$J$ models. (a) Extrapolations of binding energies versus system length for the four-leg $\sigma t$-$t'$-$J$ model at $\delta=1/8$. The binding energies are fitted smoothly to zero by a second-order polynomial function $\mathcal{C} \left(1/L_x \right)=\mathcal{C} \left(0\right)+a/L_x +b/L_x^2$. (b) and (c) show the binding energies for the $t$-$t'$-$J$ model on the $L_y = 5$ cylinder at $\delta=1/12$ ($L_x = 24, 48$) and $\delta=1/8$ ($L_x = 32, 48$), respectively. The binding energies are calculated using the energies $E(N_h, S)$ after the extrapolation to the infinite-bond-dimension limit. For both doping levels, the binding energies appear to be converged with system length $L_x$, which is similar to Figs.~\ref{fig:binding}(a) and \ref{fig:binding}(b) of the $L_y = 4$ systems. (d) The binding energies for the $t$-$t'$-$J$ model on the $L_y = 6$ cylinder at $\delta=1/8$ ($L_x = 32$). (e) and (f) show the binding energies for the $t$-$t'$-$J$ model on the $L_y = 3$ cylinder at $\delta=1/12$ ($L_x = 24, 48$) and $\delta=1/8$ ($L_x = 32, 48$), respectively. The negative binding energies demonstrate the hole pairing in these systems. Here, the binding energies of the $t$-$t'$-$J$ model at $L_y = 5, 6$ are calculated using the energies $E(N_h, S)$ after the extrapolation to the infinite-bond-dimension limit (cf. Fig.~\ref{ground_energy} and Fig.~\ref{ground_energy_5}).}
\end{figure*}

In the main text, we have shown the system length $L_x$ dependence of binding energies for the four-leg $\sigma t$-$t^{\prime}$-$J$ model at $\delta=1/12$. Here, we present the same data for $\delta=1/8$ in Fig.~\ref{fig:BE_8_extrap}(a).
The positive binding energies at finite-$L_x$ and vanished binding energy in the infinite-$L_x$ limit support the absent pairing in the $\sigma t$-$t^{\prime}$-$J$ model.

For the $t$-$t'$-$J$ model on the five-leg systems, the binding energies are negative and also well converged with system length [Figs.~\ref{fig:BE_8_extrap}(b) and \ref{fig:BE_8_extrap}(c)].
In contrast, for the five-leg $\sigma t$-$t'$-$J$ model, our DMRG calculations also find very small positive binding energies on the finite-$L_x$ cylinder (not shown here), similar to the six-leg results demonstrated in Fig.~\ref{fig:binding}(d) of the main text, which consistently indicates the absence of hole binding in the $\sigma t$-$t'$-$J$ model.
For the $L_y = 6$ $t$-$t'$-$J$ model, we also demonstrate the binding energies of $\delta = 1/8$ in Fig.~\ref{fig:BE_8_extrap}(d), which consistently are negative as well. For this case, we do not show the result at $t' = 0$ because the bond-dimension-scaling of the odd-hole energy is not very smooth, which may result in a large error bar.
As a complementary check, we have also calculated the binding energy of the three-leg cylinder [Figs.~\ref{fig:BE_8_extrap}(e) and \ref{fig:BE_8_extrap}(f)], which show a weak size dependence as well and are small negative values.

\begin{figure*}[htp]
\includegraphics[width=0.9\linewidth]{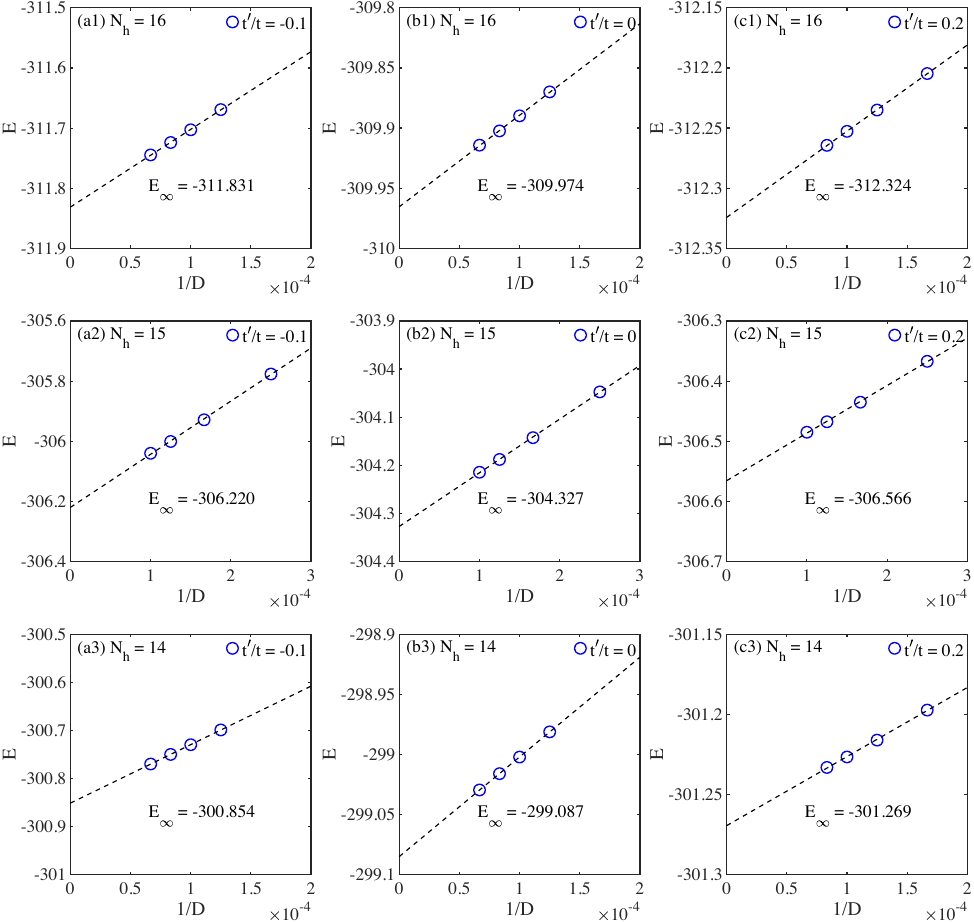}
\caption{\label{ground_energy}Extrapolations of the ground-state energies $E(N_h, S)$ for the six-leg $t$-$t'$-$J$ model on the size $L_{y} \times L_{x}=6 \times 32$. (a1-c1) show the bond dimension scaling of the ground-state energy at $t^{\prime}/t=-0.1$, $0$, and $0.2$ with the hole number $N_{h}=16$. (a2-c2) and (a3-c3) show the similar results for the hole numbers $N_{h}=15$ and $N_{h}=14$, respectively. We keep the bond dimensions $D$ up to $10000 - 15000$ SU(2) multiplets. The energies are extrapolated by a linear function $\mathcal{C} \left(1/D \right)=a/D + \mathcal{C} \left(0\right)$ to give the energy $E_{\infty}$ in the infinite bond dimension limit.}
\end{figure*}

\begin{figure*}[htp]
\includegraphics[width=0.9\linewidth]{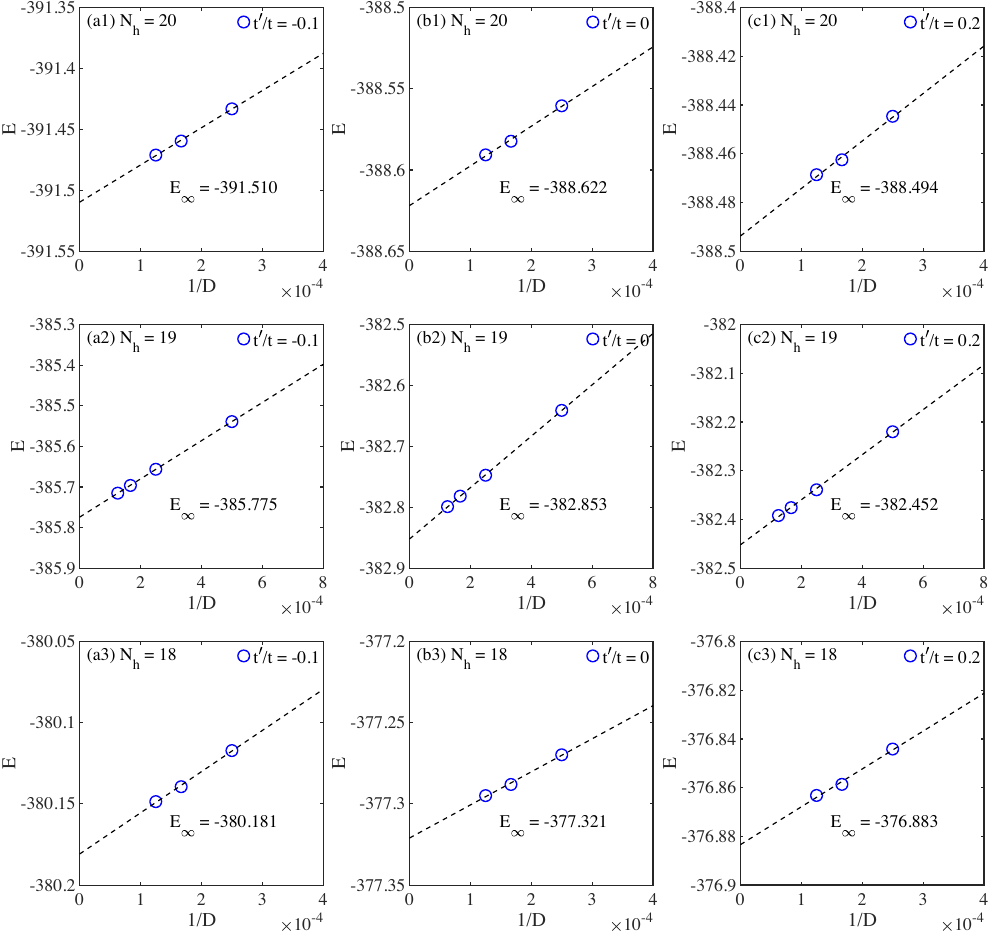}
\caption{\label{ground_energy_5}Extrapolations of the ground-state energies $E(N_h, S)$ for the five-leg $t$-$t'$-$J$ model on the size $L_{y} \times L_{x}=5 \times 48$. (a1-c1) show the bond dimension scaling of the ground-state energy at $t^{\prime}/t=-0.1$, $0$, and $0.2$ with the hole number $N_{h}=20$. (a2-c2) and (a3-c3) show the similar results for the hole numbers $N_{h}=19$ and $N_{h}=18$, respectively. We keep the bond dimensions $D$ up to $8000$ SU(2) multiplets. The energies are extrapolated by a linear function $\mathcal{C} \left(1/D \right)=a/D + \mathcal{C} \left(0\right)$ to give the energy $E_{\infty}$ in the infinite bond dimension limit.}
\end{figure*}

\begin{figure*}[t]
	\includegraphics[width=0.65\linewidth]{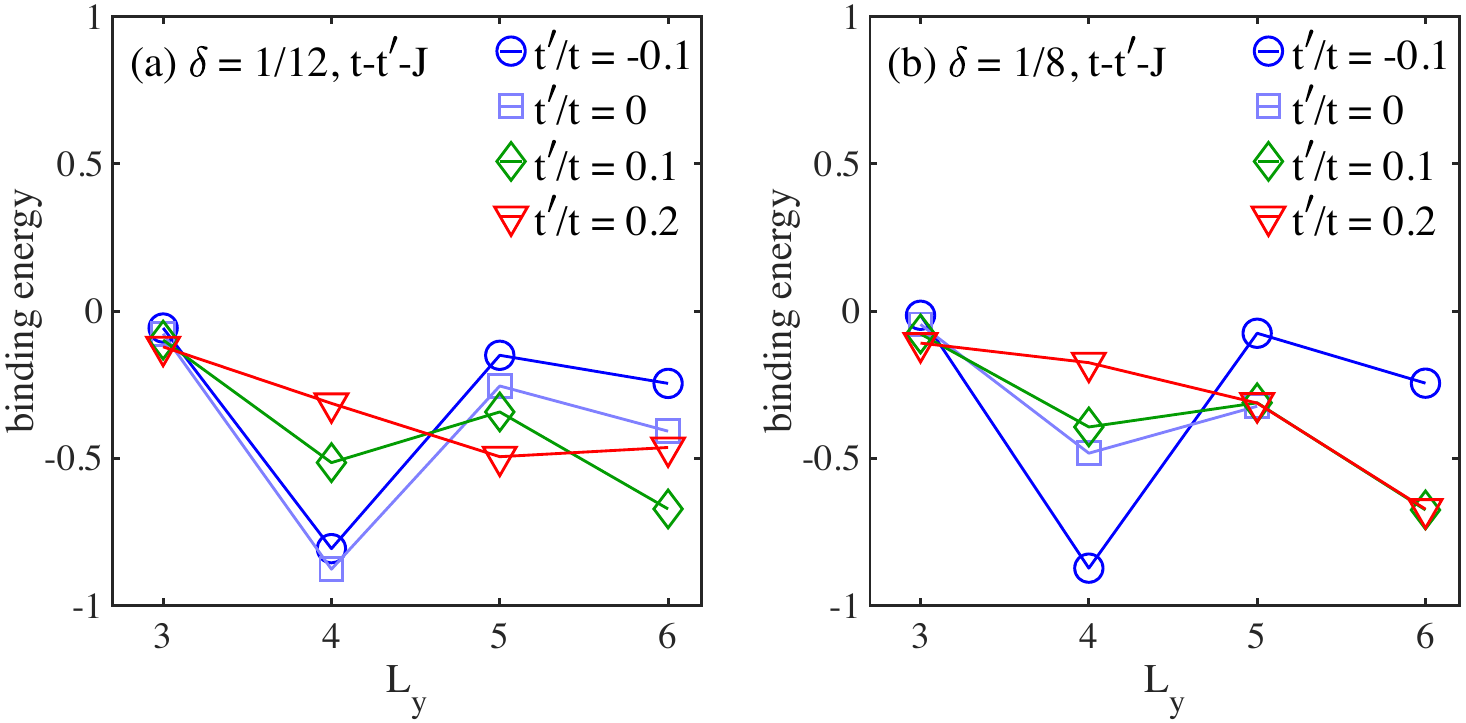}
    \caption{\label{fig:BEE}System circumference dependence of the binding energy in the $t$-$t'$-$J$ model for (a) $\delta=1/12$ and (b) $\delta=1/8$. $t'/t = -0.1, 0, 0.1$ and $0.2$. For the binding energies at $L_{y}=3,4,5$, we use the data at $L_{x}=48$. For $L_{y}=6$, we use the binding energy at $L_{x}=32$. For $\delta = 1/8$ at $L_y = 6$, we do not show the result at $t' = 0$ because the bond-dimension-scaling of the odd-hole energy for this case is not very smooth, which may lead to a large error bar.}
\end{figure*}

For calculating the binding energy on the three- and four-leg systems, we can obtain the fully converged energies. 
For the wider five- and six-leg $t$-$t'$-$J$ model we need to carefully check the convergence of the energies in the three sectors $E(N_h, S)$. 
For this purpose, we obtain the energies $E(N_h, S)$ by keeping different bond dimensions, and extrapolate these energies to the infinite-bond-dimension limit. 
Then we compute the binding energy using the extrapolated results.
In Fig.~\ref{ground_energy}, we show the bond-dimension dependence of the obtained energies $E(N_h, S)$ for $t'/t = -0.1, 0, 0.2$ on the $L_x = 32, L_y = 6$ cylinder at $1/12$ doping.
We keep the bond dimensions up to $10000 - 15000$ SU(2) multiplets to ensure a good linear extrapolation behavior of the data.
We implement the linear fitting $\mathcal{C} \left(1/D \right)=a/D + \mathcal{C} \left(0\right)$ to obtain the energy in the infinite-bond-dimension limit $E_{\infty}$, from which we calculate the binding energies shown in Fig. \ref{fig:binding}(d) of the main text. 
The similar bond-dimension-scaling of energies for $L_y = 5$ is shown in Fig.~\ref{ground_energy_5}.
Notice that for five-leg systems in Figs.~\ref{fig:BE_8_extrap}(b) and \ref{fig:BE_8_extrap}(c), we do not show the binding energies at $t'/t = -0.2$ since the energy at the $S = 1$ sector for $\delta = 1/8$ does not exhibit a smooth dependence on bond dimension, which cannot give us an accurate extrapolation of the energy versus bond dimension.

For the $\sigma t$-$t'$-$J$ model, we find that although the energies $E(N_h, S)$ get improved with increasing bond dimension, the binding energy is almost independent of bond dimension in our calculation. Therefore, we do not extrapolate the energies but use the results by keeping the largest bond dimension.

At last, we summarize the results of binding energy for the $t$-$t'$-$J$ model in Fig.~\ref{fig:BEE}.
While we cannot predict the binding energy in the $L_y \rightarrow \infty$ limit due to the limit of system size, it is found that generally the binding is stronger from the five- to six-leg system, which is a supportive signal that the hole pairing may still exist in large-size limit.

\section{DMRG measurements for three- and five-leg systems}
\label{appsec:fiveleg}

\begin{figure*}[t]
	\includegraphics[width=1.0\linewidth]{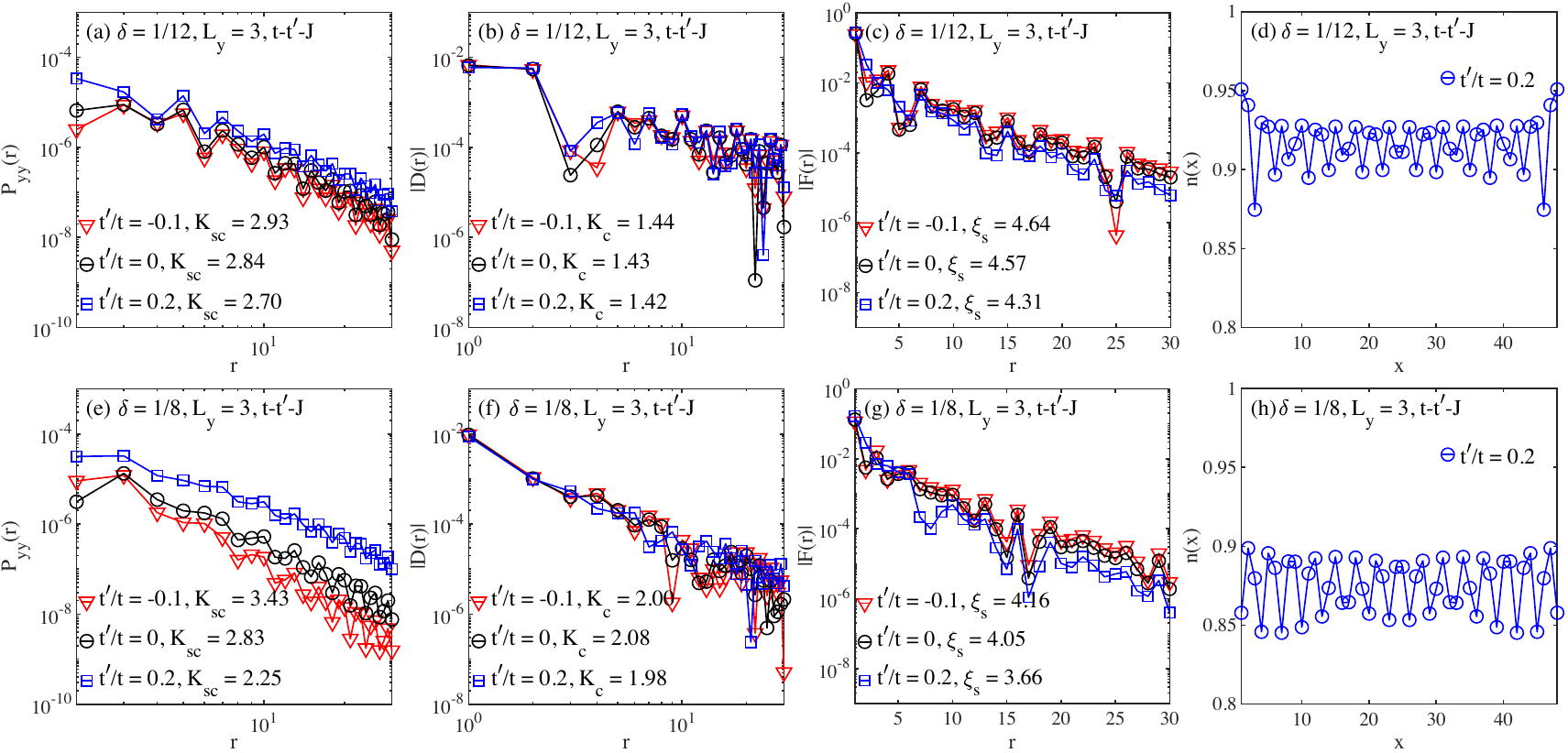}
	\caption{\label{fig:correlation3leg}Pairing correlation $P_{yy}(r)$, charge density correlation $D(r)$, spin correlation $F(r)$ and charge density profile $n(x)$ on the three-leg $t$-$t'$-$J$ model. (a-c) show the pairing correlation $P_{yy}(r)$, density correlation $D(r)$ and spin correlation $F(r)$, respectively, for different $t'/t$ on the $L_{x}=48$ cylinder with $\delta=1/12$. (d) Charge density profile $n(x)$ for $t'/t=0.2$ on the $L_{x}=48$ cylinder with $\delta=1/12$. (e)-(h) are the similar plots on the $L_{x}=48$ cylinder with $\delta=1/8$. Here, we keep $4000$ SU(2) multiplets to obtain the fully converged results. $K_{\rm sc}$ and $K_{\rm c}$ are obtained by algebraic fitting of corresponding correlation function. $\xi_{s}$ is obtained by exponential fitting of corresponding correlation function.}
\end{figure*}

\begin{figure*}[t]
	\includegraphics[width=0.65\linewidth]{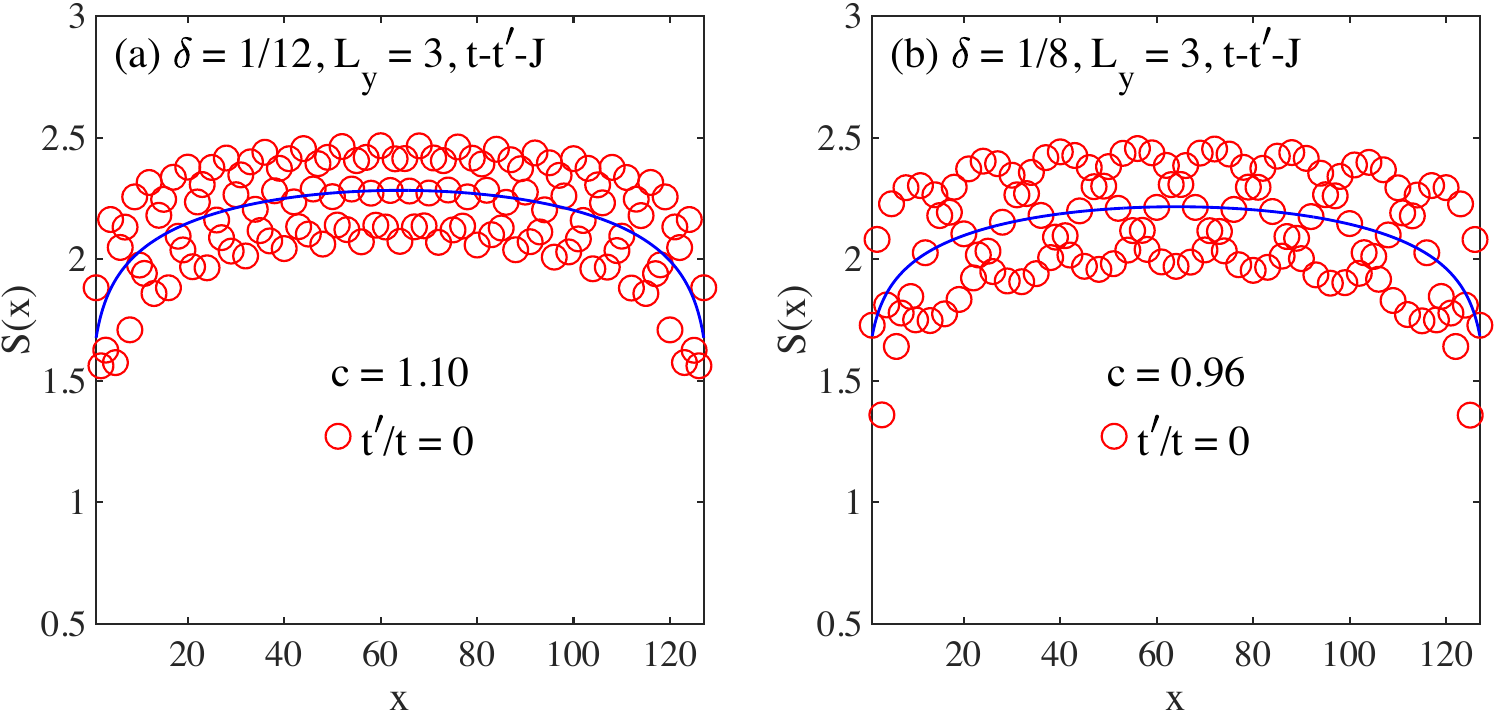}
	\caption{\label{fig:Central_charge}Fitting of central charge from entanglement entropy of the three-leg $t$-$t'$-$J$ model. (a) $t'/t=0$ on the $L_{x}=128$ cylinder at $\delta=1/12$. 
    (b) $t'/t=0$ on the $L_{x}=128$ cylinder at $\delta=1/8$. The central charge $c$ is fitted from entanglement entropy $S(x)$ by the formula $S\left(x\right)=\frac{c}{6}\mathrm{log}\left\lbrack \frac{4\left(L_x +1\right)}{\pi }\mathrm{sin}\frac{\pi \left(2x+1\right)}{2\left(L_x +1\right)}\right\rbrack +\mathrm{const.}$~\cite{CFT1,CFT2}, where $x$ denotes the column number of the subsystem in the bipartition of lattice.}
\end{figure*}

In Fig.~\ref{fig:BE_8_extrap}, we have shown the finite binding energies of the three- and five-leg $t$-$t'$-$J$ model at $\delta = 1/12, 1/8$ and $t'/t = -0.1 - 0.2$.
Here we present the corresponding measurements.
We have compared the data at different $L_x$, which give the consistent results.
For $L_y = 3$, as shown in in Fig.~\ref{fig:correlation3leg}, while both pairing correlations $P_{yy}(r)$ and density correlations $D(r)$ exhibit power-law decay, spin correlations and single-particle correlations (not shown here) follow exponential decay. 
We further determine the central charge by fitting the entanglement entropy, as demonstrated in Fig.~\ref{fig:Central_charge}, which show a good fitting giving the central charge $c \approx 1$.
Our results consistently characterize the three-leg system in the studied parameter regime as a C1S0 Luther-Emery liquid state with one gapless charge mode and no gapless spin mode, which agrees with the conclusion obtained from the DMRG simulations on the open three-leg ladder~\cite{White_1998}.

\begin{figure*}[t]
	\includegraphics[width=1.0\linewidth]{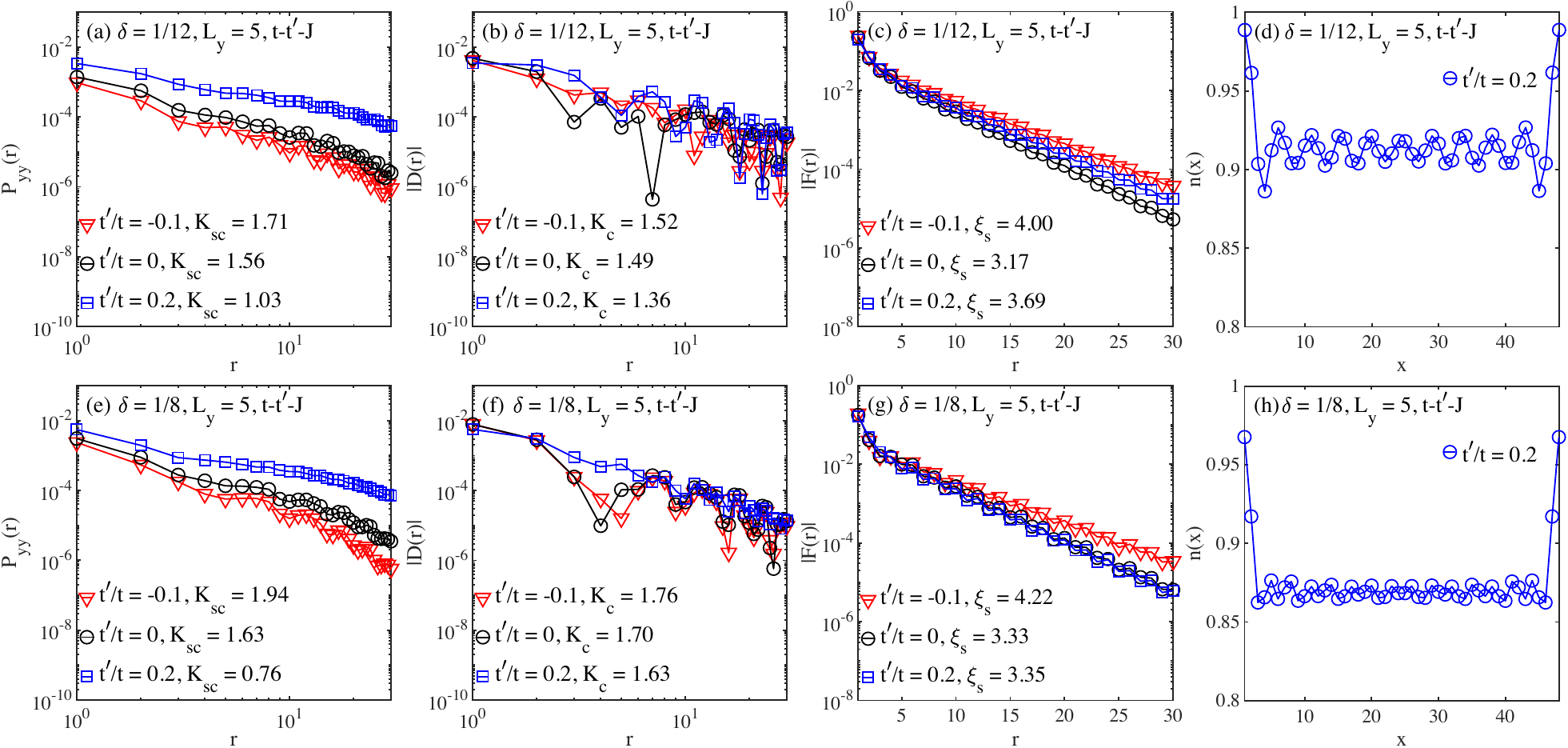}
	\caption{\label{fig:correlation5leg}Pairing correlation $P_{yy}(r)$, charge density correlation $D(r)$, spin correlation $F(r)$ and charge density profile $n(x)$ on the five-leg $t$-$t'$-$J$ model. (a-c) show the pairing correlation $P_{yy}(r)$, density correlation $D(r)$ and spin correlation $F(r)$, respectively, for different $t'/t$ on the $L_{x}=48$ cylinder with $\delta=1/12$. (d) Charge density profile $n(x)$ for $t'/t=0.2$ on the $L_{x}=48$ cylinder with $\delta=1/12$. (e)-(h) are the similar plots on the $L_{x}=48$ cylinder with $\delta=1/8$. Here, we keep $10000$ SU(2) multiplets to obtain the results. $K_{\rm sc}$ and $K_{\rm c}$ are obtained by algebraic fitting of corresponding correlation function. $\xi_{s}$ is obtained by exponential fitting of corresponding correlation function.}
\end{figure*}

For $L_y = 5$, as demonstrated in Fig.~\ref{fig:correlation5leg}, the pairing correlations can be fitted quite well as the algebraic decay with the power exponents $K_{\rm sc} < 2$, indicating the quasi-long-range superconducting order and is consistent with the hole binding shown in Fig.~\ref{fig:BE_8_extrap}.
In particular, pairing correlations are enhanced with $t'/t$ increasing from $t'/t<0$ to $t'/t>0$, which agrees with the findings on four- and six-leg systems and implies that the positive $t'/t$ can enhance the coherence of paired holes should be common on both even- and odd-leg systems.
On the other hand, the charge density correlations also appear to decay algebraically with the power exponents $K_{\rm c} < 2$, which is associated with weak charge density oscillations.
There results indicate a weak charge density order coexisting with the quasi-long-range superconducting order in the studied parameter regime of the five-leg $t$-$t'$-$J$ model.
Compared with the three-leg case in Fig.~\ref{fig:correlation3leg}, one can find the similar exponential decay of spin correlations, but pairing correlations enhance apparently at $L_y = 5$.

\section{Electron momentum distribution}
\label{appsec:nk}

\begin{figure*}[htp]
\includegraphics[width=0.6\linewidth]{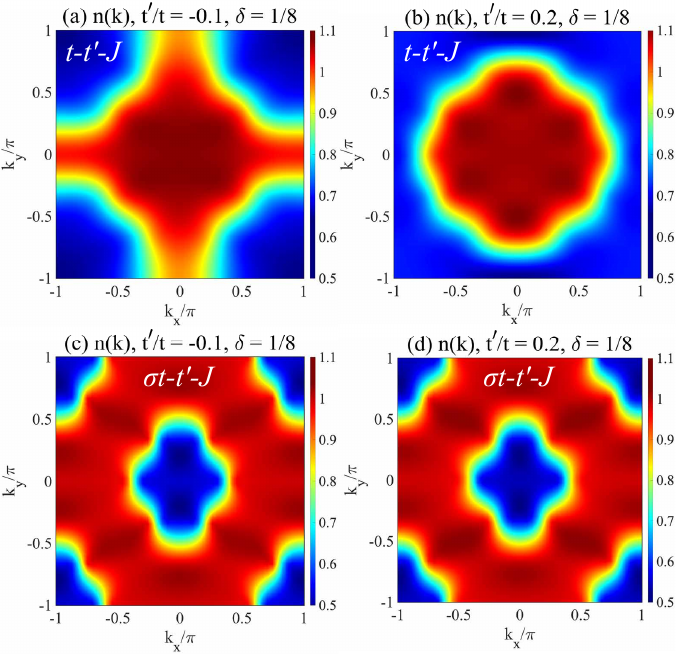}
\caption{\label{supfig:nk6-8}Electron momentum distribution $n(\bf k)$. (a) and (b) show the results of the $t$-$t'$-$J$ model, where an open and closed Fermi surface emerge in the CDW and SC phases. (c) and (d) show the results of the $\sigma t$-$t'$-$J$ model, where the electrons of spin-up and spin-down in $n(\bf k)$ are displaced by $(\pi,\pi)$. Here $L_y = 6$, $t^\prime/t = -0.1$ and $0.2$ at the doping level $\delta = 1/8$.}
\end{figure*}

\begin{figure*}[htp]
\includegraphics[width=0.6\linewidth]{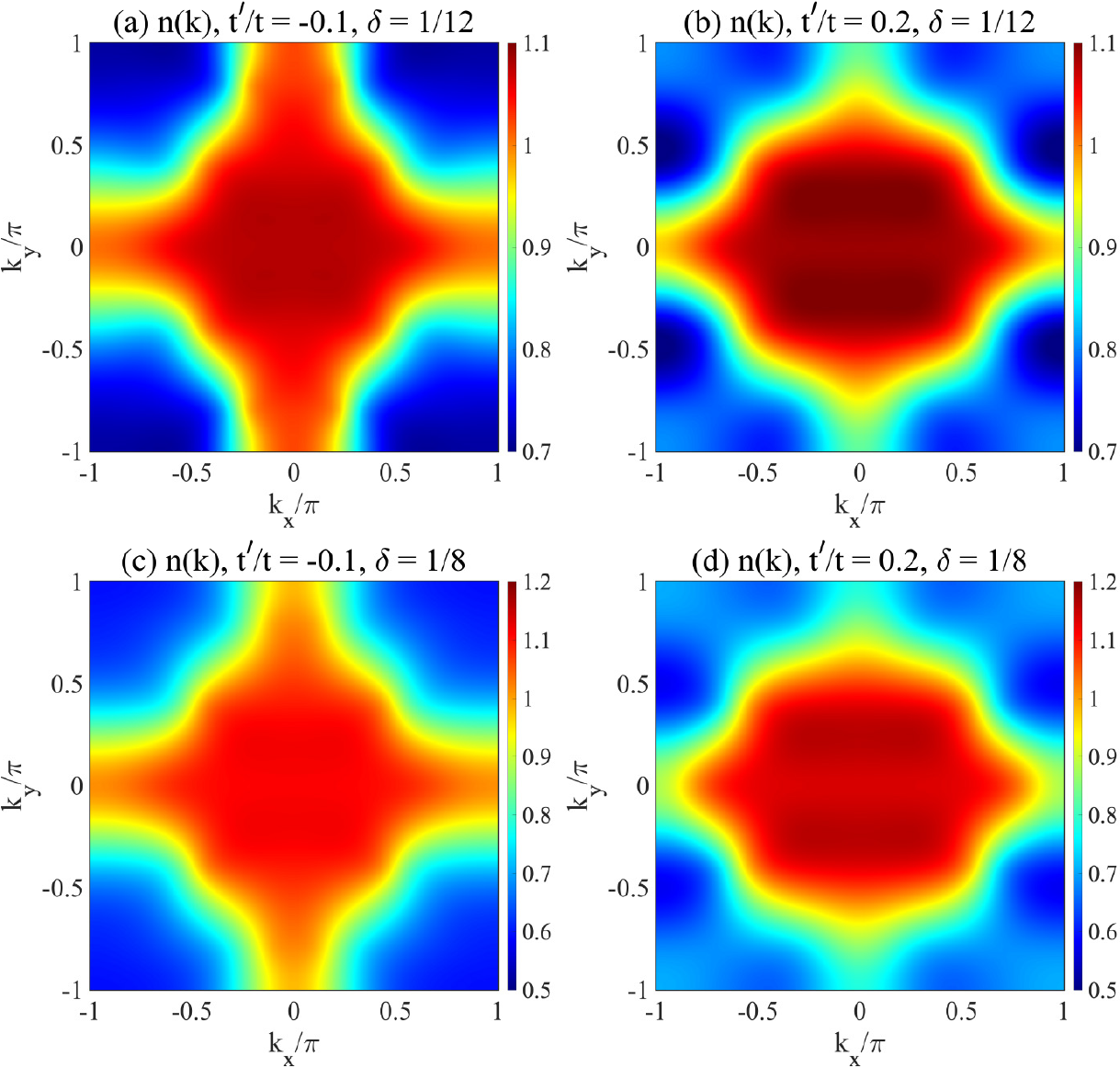}
\caption{\label{supfig:nk5leg}Electron momentum distribution $n(\bf k)$. (a) and (b) show the results of the $t$-$t'$-$J$ model at $\delta=1/12$. (c) and (d) show the similar results at $\delta=1/8$. Here $L_y = 5$, $t^\prime/t = -0.1$ and $0.2$.}
\end{figure*}

\begin{figure*}[htp]
\includegraphics[width=0.85\linewidth]{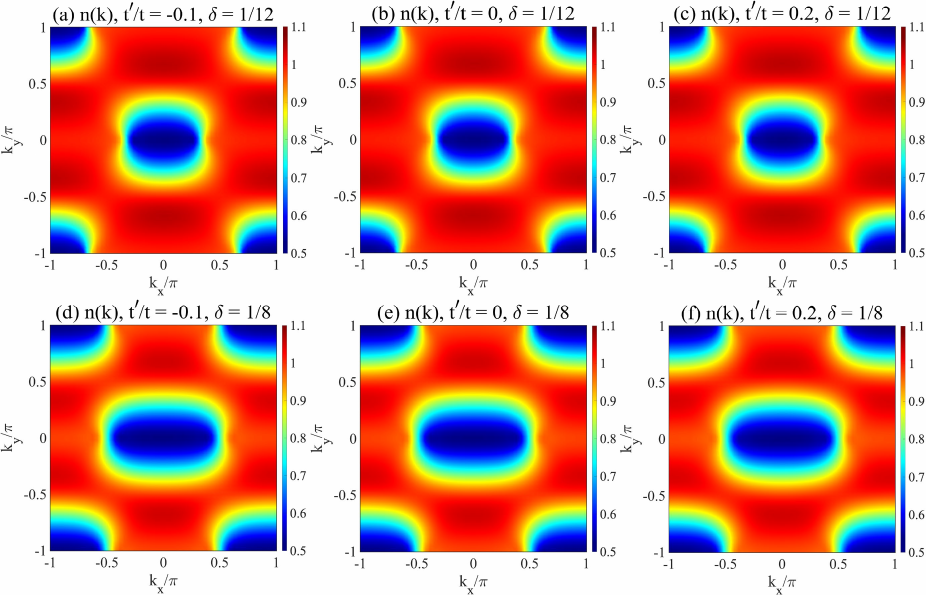}
\caption{\label{supfig:nk4}Electron momentum distribution $n(\bf k)$. (a-c) show the results of the $\sigma t$-$t'$-$J$ model at $\delta = 1/12$. (d-f) show the similar results at $\delta = 1/8$. The electrons of spin-up and spin-down in $n(\bf k)$ are displaced by $(\pi,\pi)$. Here $L_y = 4$ and $t^\prime/t = -0.1, 0, 0.2$.}
\end{figure*}

We have shown the electron momentum distributions $n\left(\mathbf{k}\right)=\frac{1}{N}\sum_{i,j,\sigma} \langle {\hat{c} }_{i,\sigma}^{\dagger } {\hat{c} }_{j,\sigma} \rangle e^{i\mathbf{k}\cdot \left({\mathbf{r}}_i -{\mathbf{r}}_j \right)}$ for the six-leg cylinder at $\delta=1/12$ in the main text. Here, we supplement with similar results for $\delta=1/8$ doping as shown in Fig.~\ref{supfig:nk6-8}. In the $t$-$t'$-$J$ model, a large Fermi surface is visible, and the topology of Fermi surface shows difference in the CDW and SC phases. By contrast, in the $\sigma t$-$t'$-$J$ model, two small Fermi pockets appear at $\mathbf{k}=\left(\pi ,\pi \right)$ and $\mathbf{k}=\left(0 ,0 \right)$, which have a weak $t'/t$ dependence and are contributed from the spin-up propagator and spin-down propagator, respectively. 
Similarly, on the five-leg $t$-$t'$-$J$ model, the topology of Fermi surface for $t'/t < 0$ and $t'/t > 0$ are distinct, as shown in Fig.~\ref{supfig:nk5leg}, which is analogous to the observations on six-leg systems [Figs.~\ref{supfig:nk6-8}(a) and \ref{supfig:nk6-8}(b)].
We also present the results for the four-leg $\sigma t$-$t'$-$J$ model in Fig.~\ref{supfig:nk4}, which show the consistent features with the results on the six-leg cylinder.

\clearpage
\twocolumngrid
\bibliography{refs}

\end{document}